\documentclass[12pt]{article}
\usepackage{amsmath}
\usepackage{amssymb}
\usepackage{graphicx,psfrag,epsf}
\usepackage{enumerate}
\usepackage{url}
\usepackage{natbib}
\usepackage{psfrag}
\usepackage{pdflscape}
\usepackage{geometry}
\usepackage{setspace}

\newtheorem{theorem}{Theorem}

\newtheorem{prop}{Proposition}

\newcommand{\blind}{0}

\begin{document}

\def\spacingset#1{\renewcommand{\baselinestretch}%
{#1}\small\normalsize} \spacingset{1}


\if0\blind
{
  \title{\bf Approximate likelihood inference in generalized linear latent variable models based on integral dimension reduction}
  \author{Silvia Bianconcini\thanks{
    The authors gratefully acknowledge the financial support from the grant RBFR12SHVV funded by the
Italian Government (FIRB project ``Mixture and latent variable models for causal inference and
analysis of socio-economic data'')}\hspace{.2cm}\\
    Department of Statistical Sciences, University of Bologna\\
    and \\
    Silvia Cagnone \\
    Department of Statistical Sciences, University of Bologna\\
    and \\
    Dimitris Rizopoulos\\
    Department of Biostatistics, Erasmus University Rotterdam.}
  \maketitle
} \fi

\if1\blind
{
  \bigskip
  \bigskip
  \bigskip
  \begin{center}
    {\LARGE\bf Title}
\end{center}
  \medskip
} \fi

\bigskip
\begin{abstract}
Latent variable models represent a useful tool for the analysis of complex data when the constructs of interest are not observable. A problem related
to these models is that the integrals involved in the likelihood function cannot be solved analytically. We propose a computational approach, referred to as Dimension Reduction Method (DRM), that consists of a dimension reduction of the multidimensional integral that makes the computation feasible in situations in which the quadrature based methods are not applicable.  We discuss the advantages of DRM compared with other existing approximation procedures in terms of both computational
feasibility of the method and asymptotic properties of the resulting estimators.\end{abstract}

\noindent%
\vfill

\newpage
\spacingset{1.45} 
\section{Introduction}

Latent variable models are commonly used in many research fields where it is of interest to study constructs that are not observable but can be indirectly measured by indicators related to them.
Typical examples can be found in sociology where questionnaires are utilized to measure attitudes and opinions, in educational testing where the ability of students is evaluated using exams, and
in medicine where quality of life is assessed by function scores (\emph{e.g.}, been able to walk independently). \par
A general framework that includes a large variety of latent variable models is represented by the Generalized Linear Latent Variable Models (GLLVMs) \citep{MouKno00,SkroRabe04}.  The purpose of
GLLVMs is to describe the relationship between a set of $p$ responses or items, denoted for each individual by the $p$-dimensional vector $\mathbf{y}_{l}, l=1, \ldots, n$, and  a
set of $q$ latent variables or/and random effects, denoted by $\mathbf{b}_{l}$, that are fewer in number than the observed variables ($q < p$). The latent variables are supposed to account for
the dependencies among the
response variables, while random effects are usually introduced in the model to account for some unobserved heterogeneity. Typically, in these models, the assumption of conditional or local
independence is made, which states that the observed response variables are independent given the latent variables. The probability associated to the individual response pattern is expressed as
follows
\begin{equation} \label{marg}
f(\mathbf{y}_{l})=\int_{\mathbb{R}^{q}}g(\mathbf{y}_{l}\mid
\mathbf{b}_{l})h(\mathbf{b}_{l})\textit{d}\mathbf{b}_{l},
\end{equation}
where $h(\mathbf{b}_{l})$ is usually referred to as the structural part of the model, generally assumed to be normal with null mean and covariance/correlation matrix equal to $\boldsymbol\Psi$,
and $g(\mathbf{y}_{l}\mid\mathbf{b}_{l})$  is usually referred to as the measurement part of the model. The latter is taken from the exponential family. For the specific case of binary items
treated in this study, assuming a canonical link function, it results
\begin{equation}\label{cond}
g(\mathbf{y}_{l}\mid \mathbf{b}_{l})= \exp \left\{\mathbf{y}^{T}_{l}\boldsymbol\eta_l- \mathbf{1}^{T} \log [ 1 + \exp(\boldsymbol\eta_l) ]\right\},
\end{equation}
where $\boldsymbol \eta_{l}$ represents the systematic component of the model defined as
\begin{equation*}
\boldsymbol \eta_{l}=\boldsymbol \alpha_{0} +\boldsymbol\alpha^{T}\mathbf{b}_{l},\
\end{equation*}
with $\boldsymbol \alpha_{0}$ being a $p$-dimensional vector of item specific intercepts and $\boldsymbol\alpha$ is a structure matrix that contains unknown factor loadings if associated to
latent variables, and fixed coefficients if associated to random effects. The measurement model can be easily extended to different kinds of observed variables. \par
GLLVMs can be estimated using either the Maximum Likelihood (ML) method, through the EM or the Newton-Raphson algorithms \citep{MouKno00,HubRonFe04}, or under the Bayesian paradigm
\citep{Dun00,Dun03}. Here, we focus on the ML estimation method. In this case, the maximization procedure requires the latent variables and/or random effects to be integrated out from the
likelihood function, but analytical solutions do not exist. Different remedies have been proposed in the literature in order to overcome this problem. Numerical quadrature based methods represent
a widespread solution to approximate the integrals. At this
regard, the Adaptive Gauss Hermite (AGH) quadrature approximation is the ``gold'' standard, especially in presence of dichotomous data \citep{HesSkPi05,SchBo05}. However, it presents two
drawbacks.
The first one is that it is very computationally demanding because of the need to find the mode of the integrand for each subject and at each iteration of the optimization algorithm.  In this
regard,  \cite{Rizo12} proposed a pseudoversion of AGH that is faster than the classical AGH since  the quadrature locations are updated just at the first iteration of the
algorithm. The second drawback is that AGH becomes computationally unfeasible in presence of a large number of latent variables.\par
An alternative way to face the problem is given by the Laplace approximation \citep{DeBru81} that can be viewed as a particular case of AGH when one quadrature point is used
\citep{LiuPie94,PiBa95}. Unlike AGH, the  Laplace approximation does not suffer from the curse of dimensionality since it does not require to solve any integral. Its computational complexity
is only related to the computation of the mode of the integrand with respect to the latent variables. The standard Laplace approximation has been applied to
GLLVMs by \cite{HubRonFe04} who investigated the properties of the resulting estimators when the direct maximization is used. In the Bayesian framework, \cite{Rue09} combined Laplace
approximation with numerical integration to provide a fast and accurate method, the so called Integrated Nested Laplace Approximation (INLA), to approximate the predictive density of the
latent variables/random effects. This approach is particularly effective when the inverse covariance matrix of the random effects is sparse and the number of parameters is small.
\par The simplicity of the standard Laplace approximation has a cost related to the fact that the order of the approximation error is $O(p^{-1})$, and it cannot be improved unless the number of
items increases. Moreover, \cite{Joe08} investigated the performance of the method
for different discrete response mixed models, and he found that it
becomes less adequate as the degree of discreteness increases. Several authors developed higher order versions of the method with the aim of improving the approximation error. A first extension
of the standard Laplace, called fully exponential Laplace approximation, was proposed  by \cite{TieKa89} in the Bayesian context,  extended by \cite{Rizo09} to joint models
for time event and longitudinal data, and applied by \cite{BiaCa12} to GLLVMs. The advantage of this method is that it improves the approximation error to
$O(p^{-2})$, but with the same computational complexity as the standard Laplace.
The good behavior of higher (than two) order Laplace approximations has been highlighted in several papers, such as \cite{ShMcCu95}, \cite{Rade00}, and \cite{Evan11}.
However, the implementation of higher order Laplace approximations requires to compute a large number of partial derivatives of the integrand, strictly
dependent
on the model considered. \par
In this paper, we propose a computational approach to approximate the integrals involved in the likelihood of GLLVMs. The proposal, that we refer to as Dimension Reduction Method (DRM), is
based on the High Dimensional Model Representation (HDMR) of continuous real functions discussed by \cite{RaAl99}. HDMR admits different specifications and the one considered here, called
Cut-HDMR \citep{Li01}, consists of representing a target function as a sum of not overlapping components that can be proved to be the terms of the Taylor series expansion of the function
\citep{XuRa04}. The application of Cut-HDMR representation truncated at $s$  ($s < q$) terms of the integrand in eq. (\ref{marg}) implies a dimension reduction of the multidimensional integral that makes the computation feasible, also when the number of latent variables is large, and, in particular, in common situations in which the quadrature based methods are not applicable. The
choice of $s$  terms involved in the expansion determines the goodness of integral approximation. It has to be noticed that, when $s$ is set equal to zero, we get the standard Laplace approximation. 
In all the other cases, the DRM method produces a reduction in the integral dimension, since only $1,2,\ldots,s$-dimensional integrals are involved.
We discuss the advantages of DRM compared with the other existing approximation procedures, that is  the Laplace approximation and  the AGH quadrature,  in terms of both  computational
feasibility of the method and asymptotic properties of the resulting estimators. As for the former, we show that, differently from the Laplace approximation methods, DRM does not require any
derivative computation, and hence is very simple to implement. As for the latter, we give a formal proof that the DRM estimators based on $s$ terms, where $s<q$, are asymptotically as accurate
as the AGH estimators. \par
The performance of the method is explored by means of a wide simulation study and on real data. In particular, we consider data coming from a longitudinal study, the Health and
Retirement Study (HRS), conducted by the University of Michigan with the aim of investigating the health, social and economic implications of the aging of the American population. We analyze the
dataset collected on a subsample of individuals who participated to a more extensive research on Aging, Demographics and Memory Study (ADAMS), which represents the first population-based study of
dementia in the United States. We first estimate a classical confirmatory factor model on the data collected at the first time of observation. We apply the proposed methodology and compare the
results with those derived by applying the standard Laplace and the AGH. Then, we highlight the main advantages of DRM with respect to the other approximation methods through a longitudinal
analysis of  the ADAMS data set observed at different occasions.\\
The paper is organized as follows. In Section 2, we illustrate the integration problem when a maximum likelihood  estimation of the model is used, and we discuss the dimension reduction
approximation of the likelihood function. In Section 3, we derive
the asymptotic properties of the DRM-based estimators and compare their behavior with that of the Laplace approximation and AGH-based estimators. The finite sample properties of the proposed
estimators are analyzed in a simulation study presented in Section 4. A comparison with the existing approximation methods, that is the standard Laplace and AGH, is also carried out. The application to the HRS data is illustrated in
Section 5 and Section 6 concludes the paper.

\section{Inference based on the Dimension Reduction Method (DRM)}\doublespacing

 Maximum Likelihood (ML) estimates in the GLLVM framework are typically obtained by using either the EM algorithm or the direct maximization through Newton-type algorithms. The key component
 for
 applying both the procedures is the score vector of the observed data log-likelihood function.
For a random sample of size $n$, the latter is defined as
\begin{footnotesize}\begin{eqnarray} \label{loglike}\ell(\boldsymbol \theta)&=&\sum_{l=1}^{n}\log f(\mathbf{y}_{l}; \boldsymbol \theta)\\\nonumber &=& -\frac{nq}{2}\log (2\pi) - \frac{n}{2}\log \boldsymbol\mid
\Psi\mid  + \sum_{l=1}^{n}\log \int_{\mathbb{R}^{q}} \exp \left\{ \mathbf{y}^{T}_{l}\boldsymbol\eta_l- \mathbf{1}^{T}\log[1+\exp(\boldsymbol\eta_l)]
- \frac{1}{2} \mathbf{b}^{T}_{l}\boldsymbol \Psi^{-1} \mathbf{b}_{l} \right\}  d\mathbf{b}_{l},\end{eqnarray}\end{footnotesize}where $\boldsymbol \theta$ denotes the vector of model parameters, that is $\boldsymbol \theta =\left[\boldsymbol \alpha_{0}, \textrm{vec}(\boldsymbol \alpha), \textrm{vec}(\boldsymbol
\Psi)\right]^{T}$.  It is easily shown that the score vector  equals
\begin{footnotesize}\begin{eqnarray} \label{score}
S(\boldsymbol \theta)=\frac{\partial \ell(\boldsymbol \theta)}{\partial \boldsymbol \theta}&=&\sum_{l=1}^{n}\frac{\partial}{\partial \boldsymbol \theta} \log f(\mathbf{y}_{l};\boldsymbol
\theta)
=\sum_{l=1}^{n}\int_{\mathbb{R}^{q}}S_{l}(\boldsymbol \theta; \mathbf{b}_{l})h(\mathbf{b}_{l} \mid \mathbf{y}_{l}; \boldsymbol \theta)d\mathbf{b}_{l}=\sum_{l=1}^{n}
E_{\mathbf{b}\mid \mathbf{y}}[S_{l}(\boldsymbol \theta; \mathbf{b}_{l})],
\end{eqnarray}\end{footnotesize}where $S_{l}(\boldsymbol \theta; \mathbf{b}_{l})$ denotes the complete-data score vector given by \linebreak $\partial \log f(\mathbf{y}_{l}, \mathbf{b}_{l}; \boldsymbol \theta)/\partial \boldsymbol
\theta=\partial \left[\log g(\mathbf{y}_{l} \mid \mathbf{b}_{l}; \boldsymbol \theta) + \log h(\mathbf{b}_{l})\right]/\partial \boldsymbol \theta$. In words, the observed data score vector is
expressed as the expected value of the complete-data vector with respect to  $h(\mathbf{b}_{l} \mid \mathbf{y}_{l}; \boldsymbol \theta)$, that is the posterior distribution of the latent
variables given the observations \citep{Lou82}. This implies that eq. (\ref{score})  plays a double role. If the score equations  are solved with respect to $\boldsymbol \theta$, with
$h(\mathbf{b}_{l} \mid \mathbf{y}_{l}; \boldsymbol \theta)$ fixed at the $\boldsymbol \theta$-value of the previous iteration, then this corresponds to the EM algorithm, whereas, if the score
equations are solved with respect to $\boldsymbol \theta$ considering $h(\mathbf{b}_{l} \mid \mathbf{y}_{l}; \boldsymbol \theta)$ also as a function of $\boldsymbol \theta$, then this
corresponds
to a direct maximization of the observed data log-likelihood $\ell(\boldsymbol \theta)$. As we shall discuss further, based on this appealing feature, the estimators derived by applying either
of
these two algorithms will share the same theoretical properties.

Eq. (\ref{score}) involves multidimensional integrals which cannot be solved analytically, except in presence of observed normal variables. An approximation of these integrals is
needed, on which the bias and variance of resulting estimators will depend.
Several numerical techniques have been proposed, but  they produce unreliable solutions or become computational unfeasible as the number of latent variables increases.
In this paper, we propose a generalized Dimension Reduction Method (DRM) developed  by \cite{RaXu04} and \cite{XuRa04}. As illustrated in the following, this method provides more accurate
estimates than the Laplace approximation and, at the same time, it avoids the main computational limitations of the adaptive quadrature.

\subsection{The generalized DRM  for multidimensional integration}

The dimension reduction method \citep{RaXu04,XuRa04} is applied to approximate the log-likelihood function (\ref{loglike}). It is based on a reduction of the dimensionality of the
integral through an additive decomposition of the $q$-dimensional integrand into at most $s$-dimensional terms, where $s <q$. This decomposition is known as Cut - High Dimensional Model
Representation (HDMR), and we refer the reader to \cite{RaAl99} for a more detailed description of the approach.\\
For unbounded integration domains, the method requires  the  integrals to
be expressed in  expected value form as follows
\begin{equation} \label{expect}
E_{h_{1}}[\nu(\mathbf{b}^{*})]=\int_{\mathbb{R}^{q}} \nu(\mathbf{b}^{*})h_{1}(\mathbf{b}^{*})d\mathbf{b}^{*},
\end{equation}
where $\nu(\mathbf{b}^{*})$ is a continuous, differentiable, real-valued function of mutually independent variables $\mathbf{b}^{*}=\left[b_{1}^{*}, \ldots , b_{q}^{*}\right]^{T}$, and $h_{1}$
is an appropriate density function. 
In order to rewrite the integral (\ref{marg}) as an expected value of the form (\ref{expect}), we follow an approach similar to the one used to derive the Laplace approximation. In this regard,
we perform the following transformation of the integrand
\begin{equation} \label{expL}
f(\mathbf{y}_{l})=\int_{\mathbb{R}^{q}}\exp [L(\mathbf{b}_{l})] d\mathbf{b}_{l}, \quad \quad l=1, \ldots , n,
\end{equation} being $L(\mathbf{b}_{l})=\log g(\mathbf{y}_{l}\mid \mathbf{b}_{l})+\log h(\mathbf{b}_{l})$. Then, we consider the Taylor expansion of $L(\mathbf{b}_{l})$ around its mode
$\widehat{\mathbf{b}_{l}}=\arg \max_{\mathbf{b}_{l} \in \mathbb{R}^{q}}L(\mathbf{b}_{l})$. That is,
\begin{equation}\label{TayL}
L(\mathbf{b}_{l})=L(\widehat{\mathbf{b}}_{l})+\frac{1}{2}(\mathbf{b}_{l}-\widehat{\mathbf{b}}_{l})^{T}L''(\widehat{\mathbf{b}}_{l})(\mathbf{b}_{l}-\widehat{\mathbf{b}}_{l})+\nu_1(\mathbf{b}_{l}),
\end{equation}
where $\nu_1(\mathbf{b}_{l})$ includes all the terms of the expansion of order higher than two. Substituting eq. (\ref{TayL}) into eq. (\ref{expL}), the integral results
\begin{eqnarray*}
\int_{\mathbb{R}^{q}}\exp [L(\mathbf{b}_{l})] d\mathbf{b}_{l} =(2\pi)^{q/2}\mid \boldsymbol{\widehat{\Sigma}}_{l}\mid ^{1/2}\exp [L(\widehat{\mathbf{b}}_{l})] \int_{\mathbb{R}^{q}} \exp[\nu_1(\mathbf{b}_{l})]
\phi(\mathbf{b}_{l};\mathbf{\widehat{b}}_{l},\widehat{\boldsymbol \Sigma}_{l})d\mathbf{b}_{l},
\end{eqnarray*}
where $\phi(\mathbf{b}_{l};\mathbf{\widehat{b}}_{l},\boldsymbol{\widehat{\Sigma}}_{l})$ represents a multivariate normal density function whose mean vector is given by the mode $\mathbf{\widehat{b}}_{l}$
and
its covariance matrix is minus the inverse of the Hessian matrix of $L(\mathbf{b}_{l})$ evaluated at its mode, that is $\boldsymbol{\widehat{\Sigma}}_{l}^{-1}=-L''(\mathbf{\widehat{b}}_{l})$. Eq.
(\ref{expL}) can be rewritten as
\begin{equation}\label{DRMlat}
f(\mathbf{y}_{l})=(2\pi)^{q/2}\mid \boldsymbol{\widehat{\Sigma}}_{l}\mid ^{1/2}\exp [L(\widehat{\mathbf{b}}_{l})]E_{\phi}\left[\exp(\nu_1(\mathbf{b}_{l}))\right].
\end{equation}
The DRM is then applied to approximate the expected value that appears in eq. (\ref{DRMlat}), after the following change of variables
\begin{eqnarray}\label{Taylor}
\int_{\mathbb{R}^{q}}\exp[\nu_1(\mathbf{\widehat{C}}_{l}\mathbf{b}_{l}^{*}+\widehat{\mathbf{b}}_{l})]\phi(\mathbf{b}_{l}^{*};\mathbf{0},\boldsymbol{I})d\mathbf{b}_{l}^{*}=\int_{\mathbb{R}^{q}}\frac{g(\mathbf{y}_{l}\mid
\mathbf{\widehat{C}}_{l}\mathbf{b}_{l}^{*}+\widehat{\mathbf{b}}_{l})h(\mathbf{\widehat{C}}_{l}\mathbf{b}_{l}^{*}+\widehat{\mathbf{b}}_{l})}{\exp\left[L(\widehat{\mathbf{b}_{l})}\right]\exp\left[-\frac{1}{2}\mathbf{b}_{l}^{*T}\mathbf{b}_{l}^{*}\right]}\phi(\mathbf{b}_{l}^{*};\mathbf{0},\boldsymbol{I})d\mathbf{b}_{l}^{*},
\end{eqnarray}
where $\mathbf{\widehat{C}}_{l}$ is obtained from the Cholesky decomposition of the covariance matrix $\boldsymbol{\widehat{\Sigma}}_{l}$, that is $\boldsymbol{\widehat{\Sigma}}_{l} =
\mathbf{\widehat{C}}_{l}\mathbf{\widehat{C}}_{l}^{T}$, and the standardized variables  $\mathbf{b}_{l}^{*}$ are derived as
$\mathbf{b}_{l}^{*}=\mathbf{\widehat{C}}_{l}^{-1}(\mathbf{b}_{l}-\mathbf{\widehat{b}}_{l})$. \\
Based on the Cut-HDMR, the function $\exp[\nu_1(\mathbf{\widehat{C}}_{l}\mathbf{b}_{l}^{*}+\widehat{\mathbf{b}}_{l})]$ in (\ref{Taylor}) is approximated by the sum of $s$ orthogonal terms as follows
\begin{footnotesize}\begin{eqnarray}\nonumber
\exp[\nu_1(\mathbf{\widehat{C}}_{l}\mathbf{b}_{l}^{*}+\widehat{\mathbf{b}}_{l})] &\approx&\sum_{i=0}^{s} (-1)^{s-i} \left(
                      \begin{array}{c}
                        q-i-1 \\
                        s-i \\
                      \end{array}
                    \right)\sum_{k_{1}<\ldots<k_{i}} \exp[\nu_{1}(\mathbf{\widehat{C}}_{l}\mathbf{b}_{k_{1}, \ldots , k_{i}}^{*}+\widehat{\mathbf{b}}_{l})]\\\label{cut2}
\end{eqnarray}\end{footnotesize}where $\mathbf{b}_{k_{1}, \ldots , k_{i}}^{*}=\left[0, \ldots , 0, b_{k_{1}}^{*}, 0, \ldots , \ldots , 0, b_{k_{i}}^{*}, 0, \ldots , 0\right]^{T}$ is a
$q$-dimensional vector characterized by $i$ free independent variables, $b_{k_{1}}^{*}, \ldots, b_{k_{i}}^{*}$, and $(q-i)$ values fixed equal to zero. Hence, the multidimensional integral
defined in expression (\ref{Taylor}) can be approximated as the sum of univariate, bivariate up to $s$-dimensional integrals as
\begin{eqnarray}\label{cut}
\lefteqn{\int_{\mathbb{R}^{q}}\exp[\nu_1(\mathbf{\widehat{C}}_{l}\mathbf{b}_{l}^{*}+\widehat{\mathbf{b}}_{l})]\phi(\mathbf{b}_{l}^{*};\mathbf{0},\boldsymbol{I})d\mathbf{b}_{l}^{*}}\\\nonumber &\approx&(-1)^{s} \left(
                      \begin{array}{c}
                        q-1 \\
                        s \\
                      \end{array}
                    \right) + \sum_{i=1}^{s} (-1)^{s-i} \left(
                      \begin{array}{c}
                        q-i-1 \\
                        s-i \\
                      \end{array}
                    \right)\sum_{k_{1}<\ldots<k_{i}} \int_{\mathbb{R}^{i}} \exp\left[-L(\widehat{\mathbf{b}_{l})}\right]g(\mathbf{y}_{l}\mid \mathbf{\widehat{C}}_{l}\mathbf{b}_{k_{1}, \ldots ,
                                       k_{i}}^{*}+\widehat{\mathbf{b}}_{l})\\\nonumber
                                       &&h(\mathbf{\widehat{C}}_{l}\mathbf{b}_{k_{1}, \ldots , k_{i}}^{*}+\widehat{\mathbf{b}}_{l})\exp\left[\frac{1}{2}\mathbf{b}_{k_{1}, \ldots ,
                                       k_{i}}^{*T}\mathbf{b}_{k_{1}, \ldots , k_{i}}^{*}\right]\phi(\mathbf{b}_{k_{1}, \ldots ,
                                       k_{i}}^{*})d\mathbf{b}_{k_{1}, \ldots , k_{i}}^{*},
                                       \end{eqnarray}
where each $i$-dimensional integral, for $i=1, \ldots, s$, can be easily computed using classical Gauss-Hermite (GH) quadratures. The approximated marginal density function results

\begin{eqnarray}\label{DRMapprox}
\tilde{f}(\mathbf{y}_{l})&=&(2\pi)^{q/2}\mid \boldsymbol{\widehat{\Sigma}}_{l}\mid ^{1/2}e^{L(\widehat{\mathbf{b}}_{l})} \left[
(-1)^{s} \left(
                      \begin{array}{c}
                        q-1 \\
                        s \\
                      \end{array}
                    \right) + \sum_{i=1}^{s} (-1)^{s-i} \left(
                      \begin{array}{c}
                        q-i-1 \\
                        s-i \\
                      \end{array}
                    \right)\right.
                                       \\\nonumber
                                       && \left. \sum_{k_{1}<\ldots<k_{i}}\sum_{t_{1}, \ldots , t_{i}} e^{-L(\widehat{\mathbf{b}}_{l})}\pi^{i/2}g(\mathbf{y}_{l}\mid \mathbf{b}_{t_{1}, \ldots ,
                                       t_{i},l})h(\mathbf{b}_{t_{1},
                                       \ldots , t_{i},l})w_{t_{1}}^{*}\ldots  w_{t_{i}}^{*}\right],
\end{eqnarray}
where $\sum_{t_{1}, \ldots , t_{i}}=\sum_{t_{1}=1}^{n_{q}} \ldots  \sum_{t_{i}=1}^{n_{q}}, i=1, \ldots, s$, being $n_{q}$  the number of quadrature points selected for each latent variable,
$\mathbf{b}_{t_{1},
\ldots , t_{i}, l}=\sqrt{2}\widehat{\mathbf{C}}_{l}(0, b_{t_{1}}, 0, \ldots , 0, b_{t_{j}},0, \ldots , 0, b_{t_{i}}, 0)^{T}+  \widehat{\mathbf{b}}_{l}$, and $w_{t_{j}}^{*}=w_{t_{j}}\exp[b_{t_{j}}^{2}]$,
where  $b_{t_{j}}$ and $w_{t_{j}}, j=1,...,i$, are the classical GH nodes and weights, respectively \citep{Stra66}.

An important feature of the Cut-HDMR (\ref{cut2}) is that each term in the summation  has a clear mathematical meaning
\citep{Li01,XuRa04}, which is especially evident if $\exp[\nu_1(\mathbf{\widehat{C}}_{l}\mathbf{b}_{l}^{*}+\widehat{\mathbf{b}}_{l})]$ can be expanded as a convergent Taylor series around $\mathbf{0}$.
Indeed, it is easy to prove that the first term, obtained for $i=0$, is the constant term of the Taylor series, the second term, derived for $i=1$, is the sum of all the Taylor series terms
which
only contain the variable $b_{k}^{*}, k=1, \ldots, q$, while, when $i=2$,  the term is the sum of all the Taylor series terms which only contain variables ${b}_{k_{1}}^{*}, k_{1}=1, \ldots, q$,
and ${b}_{k_{2}}^{*}, k_{2}=1, \ldots, q$,  and so on. In particular, \cite{XuRa04} proved the following result:
\begin{theorem}\label{Th1}
For any $q$-dimensional function $\nu(\mathbf{b}^{*})$, if it admits an $s$-variate Cut-HDMR  of the form  (\ref{cut2}), then this
approximation, denoted by $\widehat{\nu}_{s}$, consists of all
terms of the Taylor series of  $\nu(\mathbf{b}^{*})$ that have less than or equal to $s$ variables, \emph{i.e.}$$\widehat{\nu}_{s}=\sum_{w=0}^s
t_{w},$$ where

\begin{equation}\nonumber 
t_{w}=\sum_{j_{1},j_{2},\ldots,j_{w}=1}^{\infty}\sum_{k_{1}<k_{2}<\ldots<k_{w}}\frac{1}{j_{1}!j_{2}!\ldots j_{w}!}\left.\frac{\partial
^{j_{1}+j_{2},\ldots,+j_{w}} \nu(\mathbf{b}^{*})}{\partial
b_{{k_{1}}}^{*j_{1}}\partial b_{{k_{2}}}^{*{j_{2}}}\ldots\partial b_{{k_{w}}}^{*j_{w}}}\right|_{\mathbf{b}^{*}=\mathbf{0}}  b_{{k_{1}}}^{*j_{1}}b_{{k_{2}}}^{*j_{2}}\ldots
b_{{k_{w}}}^{*j_{w}},
\end{equation}
\end{theorem}

It has to be noticed that the approximated Cut-HDMR expansion provides a better approximation and convergent solution of
$\exp[\nu_1(\mathbf{\widehat{C}}_{l}\mathbf{b}_{l}^{*}+\widehat{\mathbf{b}}_{l})]$
than any truncated Taylor series expansion that
contains a finite number of terms, generally of first and second order. Indeed, from eq. (\ref{DRMapprox}), it can be noticed that, when $s=0$, the quantity
among squared brackets reduces to one, such that the DRM approximated function results
\begin{equation}\label{laplace}\tilde{f}(\mathbf{y}_{l})=(2\pi)^{q/2}\mid \boldsymbol{\widehat{\Sigma}}_{l}\mid ^{1/2}e^{L(\widehat{\mathbf{b}}_{l})},\end{equation}
that is the Laplace approximation of the integral (\ref{marg}). On the other hand, when the integral is $q$-dimensional, that is no dimensional reduction is performed, the  approximation results
\begin{equation}\label{AGH}\tilde{f}(\mathbf{y}_{l})=2^{\frac{q}{2}}\mid \widehat{\boldsymbol \Sigma}_{l}\mid ^{1/2} \sum_{t_{1}, \ldots , t_{q}} g(\mathbf{y}_{l}\mid \mathbf{b}^{*}_{t_{1}, \ldots ,
t_{q},l})h(\mathbf{b}^{*}_{t_{1}, \ldots , t_{q},l})w_{t_{1}}^{*}\ldots  w_{t_{q}}^{*},\end{equation}
where $\mathbf{b}^{*}_{t_{1}, \ldots , t_{q},l}=(b_{t_{1},l}^{*}, \ldots , b_{t_{q},l}^{*})^{T}= \sqrt{2}\mathbf{C}_{l}(b_{t_{1}}, \ldots , b_{t_{q}})^{T}+ \widehat{\mathbf{b}}_{l}$ and
$w_{t_{j}}^{*}=w_{t_{j}}\exp[b_{t_{j}}^{2}]$
are the adaptive Gauss-Hermite nodes and weights, respectively. 
In other words,  the dimension reduction method provides an approximation of the marginal density function $f(\mathbf{y}_{l})$ that is more accurate than the one derived by applying the
classical
Laplace approximation, due to the inclusion of  higher (than two) order terms in the Taylor series expansion of $L(\mathbf{b}_{l})$. The adaptive Gauss-Hermite approximation (\ref{AGH}) is
obtained when no reduction is performed, that is when the integral dimension is equal to the number of latent factors. As shown in the subsequent sections, the selection of a value of $s$ less than $q$  in the DRM
approximation avoids the main computational limitations of the adaptive quadrature, preserving, asymptotically, the same accuracy of the estimates. \\
Based on  (\ref{DRMapprox}), the approximation of the log-likelihood function becomes

\begin{eqnarray}\label{approLik}
\tilde{\ell}(\boldsymbol \theta)&=&\sum_{l=1}^{n} \log \tilde{f}(\mathbf{y}_{l}; \boldsymbol \theta)\\\nonumber
&=&\frac{nq}{2}\log(2\pi)+ \frac{n}{2}\mid \boldsymbol{\widehat{\Sigma}_l}\mid  + \sum_{l=1}^{n} \log \left[
 (-1)^{s} \left(
                      \begin{array}{c}
                        q-1 \\
                        s \\
                      \end{array}
                    \right)+\sum_{i=1}^{s} (-1)^{s-i} \left(
                      \begin{array}{c}
                        q-i-1 \\
                        s-i \\
                      \end{array}
                    \right)\right.
                                       \\\nonumber
                                       && \left. \sum_{k_{1}<\ldots<k_{i}}\sum_{t_{1}, \ldots , t_{i}} \pi^{i/2}g(\mathbf{y}_{l}\mid \mathbf{b}_{t_{1}, \ldots , t_{i},l})h(\mathbf{b}_{t_{1},
                                       \ldots ,
                                       t_{i},l})w_{t_{1}}^{*}\ldots  w_{t_{i}}^{*}\right].
\end{eqnarray}
In this study, the maximization of expression (\ref{approLik}) is performed using a quasi-Newton algorithm in which the gradient and
the Hessian matrix are obtained using numerical derivatives.

\section{Statistical properties of the DRM-based estimators}
Motivated by the real data application, we derive the properties of the DRM-based estimators in GLLVM for binary data, where $q$ latent factors $\mathbf{b}_{l}$ account for the
associations among the observed variables.\\ To investigate the properties of the DRM-based maximum likelihood estimators $\widehat{\boldsymbol \theta}$, the asymptotic behavior of the
approximation (\ref{approLik}) is analyzed by
 taking into account for its equivalent Taylor series expansion provided in Theorem \ref{Th1}.  Our arguments are similar to those of \cite{LiuPie94} and \cite{Bia14} who analyzed the
 consistency
 of the adaptive Gauss-Hermite based estimators in mixed models and GLLVM, respectively.
A detailed derivation of the error rate of the approximation is given in Appendix A.
\begin{prop}[\textbf{Consistency}]\label{cons}  Let $\boldsymbol \theta_{0} \in \Theta$ denote the true parameter value, then, under suitable regularity conditions,
\begin{equation}\label{consistency}
(\widehat{\boldsymbol \theta} - \boldsymbol \theta_{0})=O_{p}\left[\max\left(n^{-1/2},p^{-\left[\frac{n_{q}}{3}+1\right]}\right)\right].
\end{equation}
\end{prop}
\noindent Thus, $\widehat{\boldsymbol \theta}$ is consistent as long as both $n$ and $p$ grow to $\infty$. A formal proof of Proposition \ref{cons} is given in Appendix B. The $n^{-1/2}$ term comes
from the standard asymptotic theory, whereas the $p^{-\left[\frac{n_{q}}{3}+1\right]}$ term derives from the DRM approximation.
For $n_{q} \geq 3$, the DRM-based estimators are more accurate than the classical Laplace ones, that are of order $O(p^{-1})$. In particular, the DRM-based estimators share the same accuracy of
the AGH ones, but the dimension reduction method avoids the main computational limitations of the latter.\\
As discussed by \cite{Hub09} and \cite{Bia14} for the Laplace-based and AGH-based estimators, respectively, when numerical techniques are used, such as the DRM, the corresponding estimators are not maximum likelihood estimators because of the approximation, but belong to the class of $M$-estimators. Hence,
\begin{prop}[\textbf{Asymptotic normality}]\label{asynorm} If $\boldsymbol 	\theta_{0}$ is an interior point of the parameter space $\Theta$ and $H(\boldsymbol \theta_{0})=-E\left[\frac{\partial^{2} \tilde{\ell}(\boldsymbol \theta_{0})}{\partial \boldsymbol \theta\partial \boldsymbol \theta^{T}}\right]$ is nonsingular, the DRM based estimators are asymptotically normal, \emph{i.e.}
\begin{equation}\label{Mest}\sqrt{n}(\hat{\boldsymbol \theta} - \boldsymbol \theta_{0}) {\rightarrow}^{D} MVN(\mathbf{0}, H(\boldsymbol \theta_{0})^{-1}U(\boldsymbol \theta_{0})[H(\boldsymbol \theta_{0})^{-1}]^{T})\end{equation}
 with $U(\boldsymbol \theta_{0})=E\left[\frac{\partial \tilde{\ell}(\boldsymbol \theta_{0})}{\partial \boldsymbol \theta}{\frac{\partial \tilde{\ell}(\boldsymbol \theta_{0})}{\partial \boldsymbol \theta}}^{T}\right]$, being $\tilde{\ell}(\boldsymbol \theta_{0})$ given in eq. (\ref{approLik}).
\end{prop}

\section{Finite sample results}
The performance of the proposed method  is evaluated in finite samples by means of a Monte Carlo simulation study for the GLLVM in presence of binary data. This is a particular case of the model
described in
Section 1, where we consider $\mathbf{b}_{l}=(z_{1l},\ldots,z_{ql})^{T}$ as a vector of $q$ factors that account for the associations among the observed variables, and we assume
$\boldsymbol\Psi$
to be a
correlation matrix for identifiability reasons explained below.\par We consider two different scenarios corresponding to models with two and four correlated latent variables, respectively. The
choice of a reduced number of latent variables in the first scenario  allows us to compare the performance of DRM with other approximation techniques without incurring in computational problems
that, as known, can affect the quadrature-based methods when the dimension of integrals is high. In both scenarios, the sample size is set to 500 individuals. $100$ replications are
considered per each
condition of the study.

In the first scenario, the aim is to compare the performance of the DRM method with $s=1$, that means we approximate the two-dimensional integral as a sum of univariate integrals, with the
classical Laplace approximation, that corresponds to $s=0$, and with the AGH quadrature-based method, that occurs when the approximated integral is of the same dimension as the number of latent variables, in this case two. 
For both DRM and AGH, an increasing number of quadrature points is considered, that is $n_{q}$ equal to 5, 7, and 11. It has to be noticed that for this particular model AGH is
feasible for all the selected number of quadrature points.\\ The true values of the model parameters have been selected so that at least four ($q^2$) constraints are imposed to obtain  model
identification \citep{Jore69}. In particular, two constraints are obtained from the correlation matrix $\Psi$, being the main diagonal elements equal to one. Fixing the variances of the latent
variables equal to the unity also ensures the identification of their scale.
The free correlation parameter $\psi_{12}$ is set equal to $0.469$.  The remaining constraints are obtained by assuming a simple structure for the matrix of the loadings, that is by
partitioning the manifest variables into two non-overlapping groups, each indicative of a different latent variable. The true values of the free loadings are generated from a log-normal
distribution and are equal to $\alpha_{11}=2.697,\alpha_{12}=0.933,\alpha_{23}=1.232$, and $\alpha_{24}=1.634$. Starting values have been randomly chosen in a suitable range for the parameters.

Table \ref{tab01} reports the estimates of the two-factor model parameters derived by applying the Laplace approximation, DRM and AGH, the latter both based on five quadrature points
($n_{q}=5$),
and denoted by  DRM5 and AGH5, respectively. The results are illustrated in terms of Relative bias (Rbias) and Standard deviations (S.d.) of the Monte Carlo estimates, as well as in terms of
computational performance of the algorithms. As for the latter, we can observe that the percentage of samples for which the algorithm reaches convergence properly (\textit{\% cv}) increases with
$s$,  varying from $91\%$ under the Laplace approximation to $100\%$ for AGH5. The average log-likelihood values over the one hundred replications increase with $s$, whereas the average number
of
function evaluations (\emph{nr feval}) and the average computational time in seconds (\textit{avg time}) are the lowest when the  Laplace approximation is applied  and increase with $s$.  This
is clearly due to the increasing dimension of the integrals to be approximated, being unidimensional when the DRM5 is applied and bidimensional for the AGH5. In terms of accuracy of the
estimates, we can observe that the relative bias of the estimates derived by applying the Laplace approximation is quite high for almost all the loadings and for the correlation estimate,
whereas they generally improve  for increasing values of $s$. As for the standard deviation of the estimates, considering AGH5 as the ``gold" standard, we can observe that the Laplace approximation tends to strongly underestimate the variability of the loading estimates, whereas it overestimates the variability of the correlation estimate.  A similar behavior is observed for the standard deviation of the estimates under DRM5, even if they are closer to the ones obtained with AGH5

\begin{figure}[!htbp]
\centering
\psfrag{a11}{$\alpha_{11}$}\psfrag{a12}{$\alpha_{12}$}\psfrag{a21}{$\alpha_{23}$}\psfrag{a22}{$\alpha_{24}$}
\psfrag{corr12}{$\psi_{12}$}\psfrag{Estimates}{\tiny estimates}\psfrag{Density}{\tiny density}
\makebox{\includegraphics[height=5.5cm,width=5.5cm]{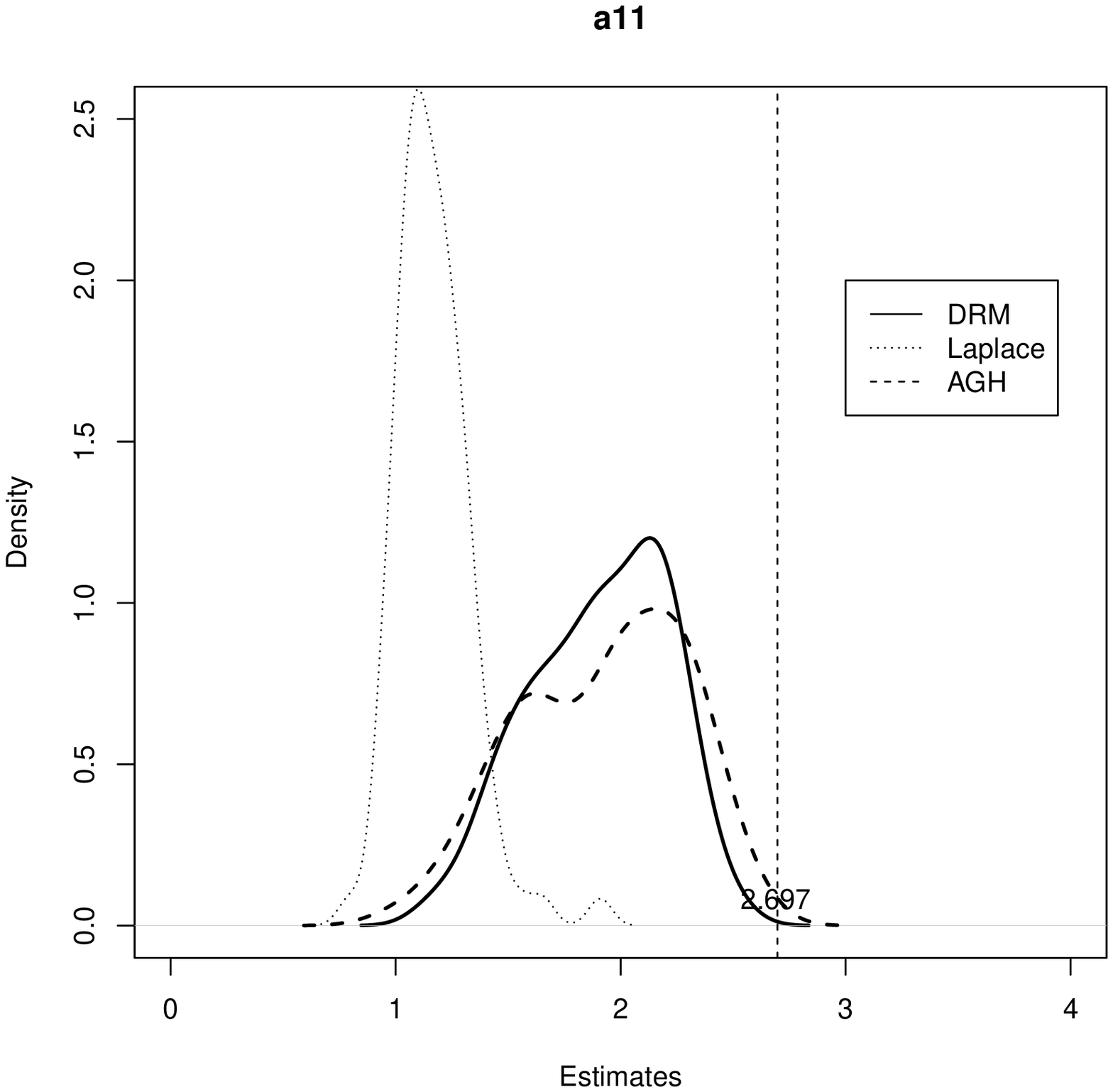} \includegraphics[height=5.5cm,width=5.5cm]{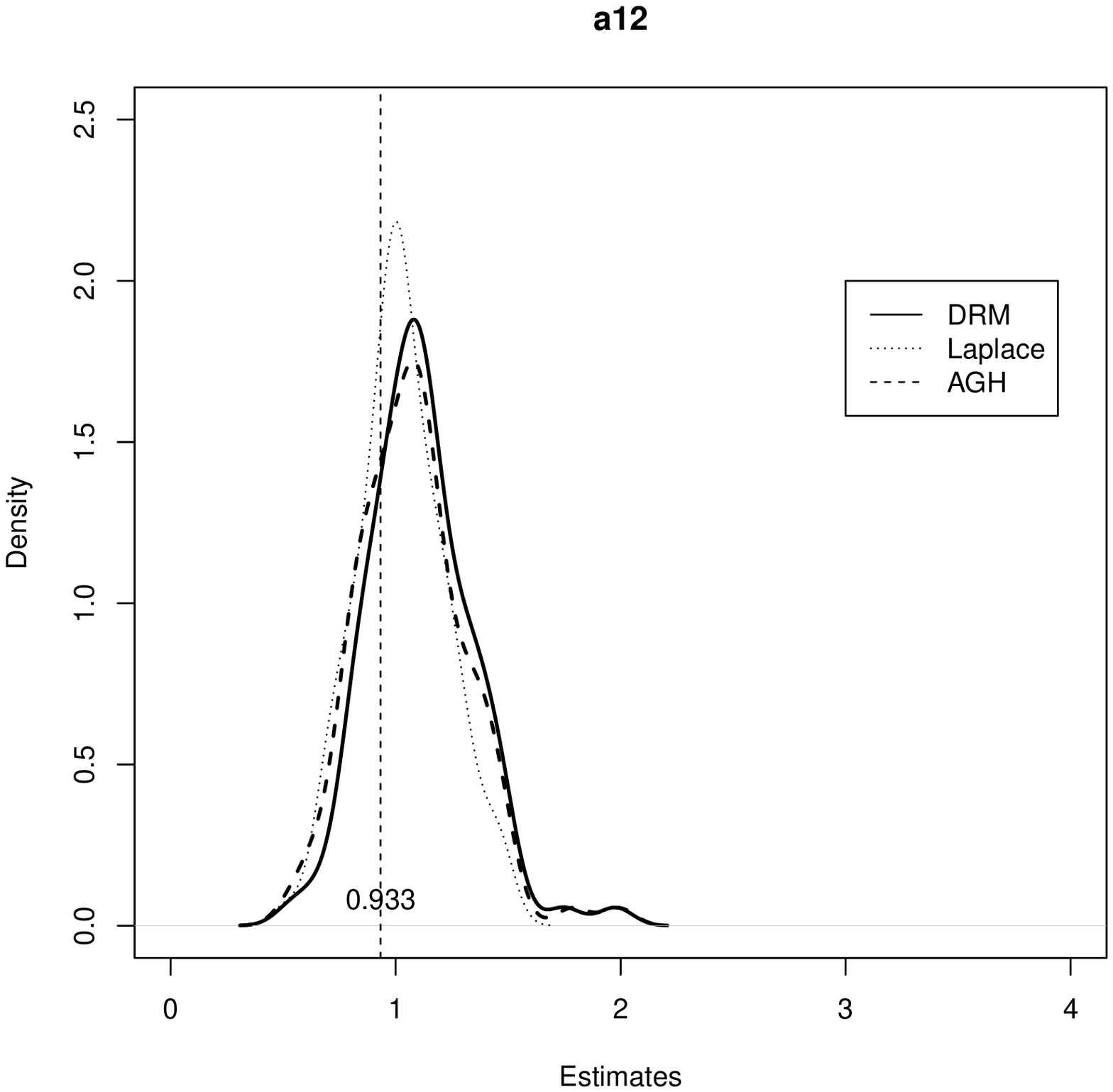}}
\makebox{\includegraphics[height=5.5cm,width=5.5cm]{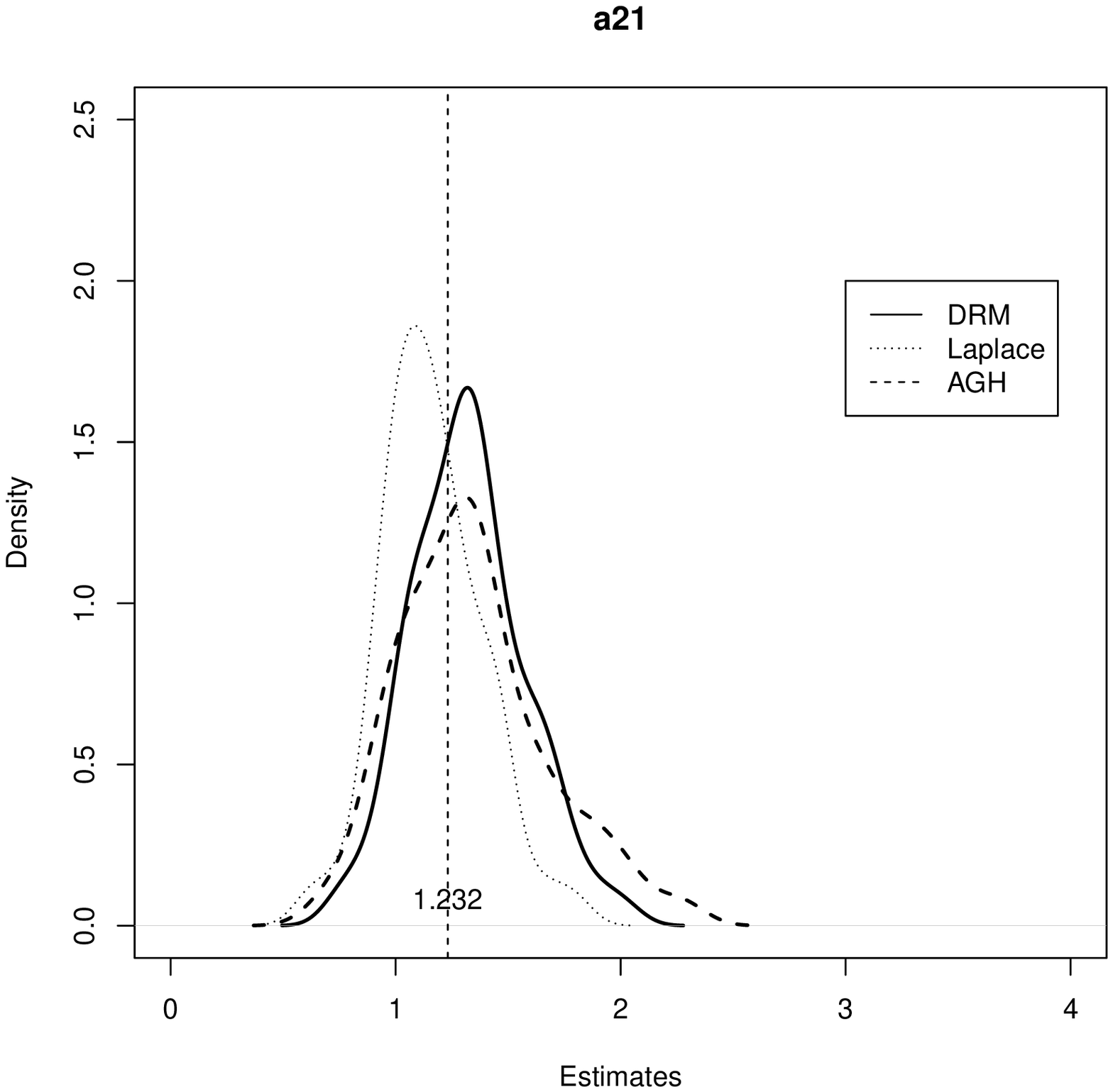} \includegraphics[height=5.5cm,width=5.5cm]{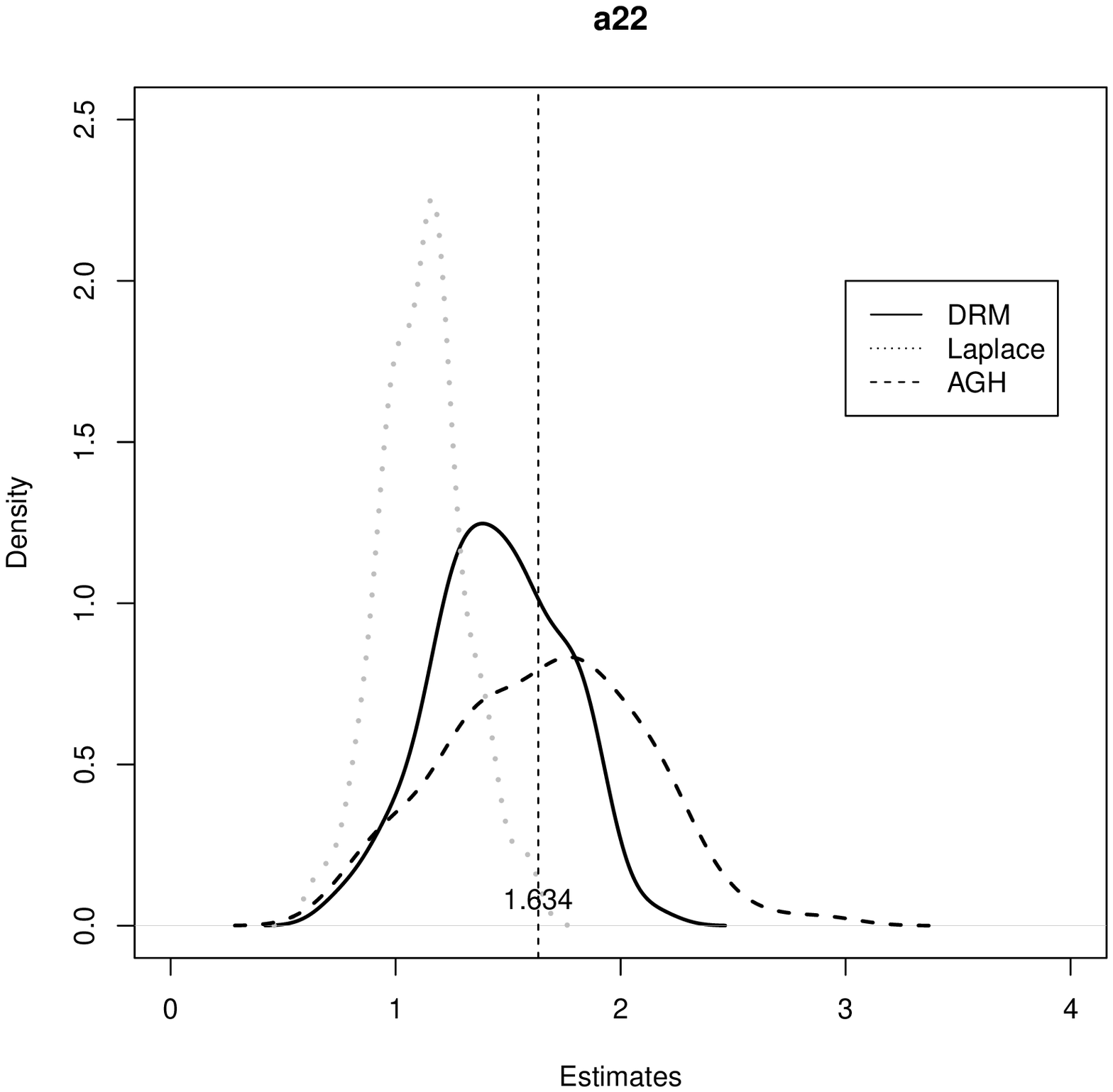}}
\makebox{\includegraphics[height=5.5cm,width=5.5cm]{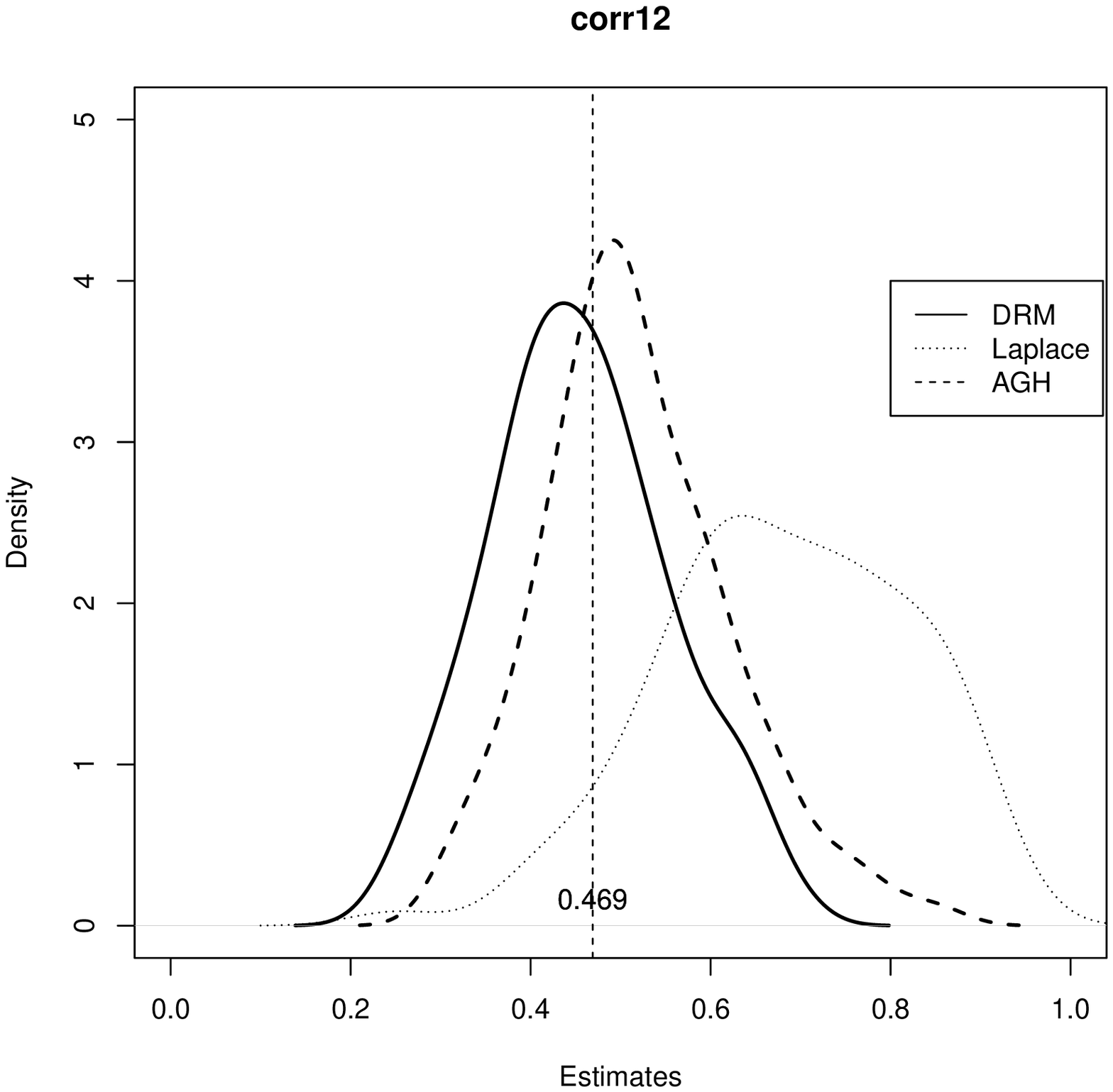}}
\caption{\label{fig01} Estimated densities of the loadings and the correlation estimates for the two-factor model under Laplace, DRM5 and AGH5. The dashed vertical line indicates the true value
of the parameter.}
\end{figure}

The behavior of the estimators obtained under the different approximations is highlighted in Figure \ref{fig01}, where their estimated densities are illustrated. It is clearly evident the different performance of the methods particularly for $\alpha_{11}$, $\alpha_{24}$, and for $\psi_{12}$. The better performance of the most informative AGH5  compared to the other
methods is particularly evident in the reduction of the bias. For the correlation parameter, the accuracy of Laplace is noticeably worse than the
other methods.
DRM5  seems to be a good compromise between the Laplace approximation and the AGH5 in terms of both accuracy of the estimates and computational burden of the algorithm.\\

\begin{center}
\begin{table}
\caption{\label{tab01}Monte Carlo results for the two-factor model under Laplace, DRM and AGH with $n_{q}=5$, $n=500$, 100 replications}\footnotesize \singlespacing
\centering
\begin{tabular}{lrrrrrrrrrrrr} \hline\tiny
&\multicolumn{3}{c}{\emph{Laplace}} &&\multicolumn{3}{c}{\emph{DRM5}}&&\multicolumn{3}{c}{\emph{AGH5}} \\\hline
&\multicolumn{3}{c}{\emph{s=0}} &&\multicolumn{3}{c}{\emph{s=1}}&&\multicolumn{3}{c}{} \\
 True                 &  Mean  & RBias   & S.d.  &&  Mean &  RBias & S.d.  && Mean &  RBias & S.d. \\		
$\alpha_{11}= 2.697$& 1.168 &  -0.567 & 0.165 && 1.908& -0.293 & 0.303 && 1.923 &-0.287 & 0.471   \\
$\alpha_{12}= 0.933$& 1.027 &  0.101 &  0.195  && 1.110 & 0.190 & 0.231 && 1.075 &0.153 & 0.242          \\
$\alpha_{23}= 1.232$& 1.161 &  -0.058 & 0.221 && 1.321& 0.072 & 0.250 && 1.342 &0.089 & 0.384       \\
$\alpha_{24}= 1.634$& 1.121 &  -0.314 & 0.189 && 1.454& -0.110 & 0.288 && 1.661 &0.023 & 0.478        \\
$\psi_{12}= 0.469$&   0.682 &	0.453  &  0.138 && 0.455&-0.030 & 0.099 && 0.519 & 0.105&  0.049         \\
\hline
\emph{Avlog-lik} &\multicolumn{3}{c}{-1225.87}&&\multicolumn{3}{c}{-1216.95}&&\multicolumn{3}{c}{-1213.95}\\
\emph{\%cv}&\multicolumn{3}{c}{91}&&\multicolumn{3}{c}{93}&&\multicolumn{3}{c}{100}\\
\emph{Nrfeval}& \multicolumn{3}{c}{16.57}&& \multicolumn{3}{c}{24.37} &&\multicolumn{3}{c}{31.17}	\\
\emph{Avtime (s)} &\multicolumn{3}{c}{38.82}&&\multicolumn{3}{c}{66.78}&&\multicolumn{3}{c}{87.71}\\\hline
\end{tabular}
\end{table}\end{center}
In Table \ref{tab012}, the results of the two-factor model estimation based on the dimension reduction method and the AGH quadrature are compared when seven and eleven quadrature
points are used
in both the numerical techniques.  DRM7 and AGH7 denote the two techniques when seven quadrature points are used ($n_{q}=7$), whereas DRM11 and AGH11 indicate the techniques when eleven
nodes are considered ($n_{q}=11$). For both seven and eleven nodes, DRM appears to be superior to the AGH in terms of average number of function evaluations and  computational time. As in the previous case, the standard deviations of the loading estimates obtained with DRM7 are smaller than those obtained with AGH7, and those obtained with DRM11 are smaller than those obtained with AGH11. For the correlation estimate DRM overestimates its standard deviation compared to AGH in both cases, $n_{q}=7$ and $n_{q}=11$. 

 \newpage
In the second scenario, we evaluate the performance of DRM for all possible values of $s$, that is zero, one, two, three. This allows to compare, also in this case, the
results of DRM with Laplace ($s=0$) and AGH. The latter is feasible only if we set the number of nodes for each dimension equal to five or seven. We consider the solution for $n_q=5$
since there are no relevant differences with $n_q=7$. Also in
this scenario, we consider a simple structure for the matrix of the loadings that is characterized by two free parameters per each factor. In particular, the loading parameters are fixed to
$\alpha_{11}=2.697, \alpha_{12}=0.933, \alpha_{23}=1.659, \alpha_{24=}1.241, \alpha_{35}=1.486, \alpha_{36}=1.156, \alpha_{47}=0.756,\alpha_{48}=0.884$  and the correlation parameters are fixed
to $\psi_{12}=0.470,\psi_{13}=0.534, \psi_{14}=0.480,\psi_{23}=0.405,\psi_{24}=0.440,\psi_{34}=0.571.$

\begin{table}\tiny
\hspace{-1cm}\caption{\label{tab012}Monte Carlo results for the two-factor model under DRM and AGH with $n_{q}=7,11$, $n=500$, 100 replications}
\hspace{-1cm}\centering %
\begin{tabular}{lrrrrrrrrrrrr} \hline
&&&&&&&&&&\\
&\multicolumn{3}{c}{\emph{DRM7}}&\multicolumn{3}{c}{\emph{AGH7}} &\multicolumn{3}{c}{\emph{DRM11}}&\multicolumn{3}{c}{\emph{AGH11}} \\\hline
&\multicolumn{3}{c}{\emph{s=1}} &\multicolumn{3}{c}{}&\multicolumn{3}{c}{\emph{s=1}}&\multicolumn{3}{c}{} \\
\cline{2-4} \cline{5-7} \cline{8-10}\cline{11-13} \\
 True                 &  Mean  & RBias   & S.d.  &  Mean &  RBias & S.d.  & Mean &  RBias & S.d. & Mean &  RBias & S.d.\\\\
$\alpha_{11}= 2.697$& 2.100&-0.222&0.457 & 2.214 &-0.179& 0.553  & 2.525  & -0.064& 0.899& 2.995 &0.110 & 1.187 \\
$\alpha_{12}= 0.933$& 1.088&0.166&0.275& 1.044   &0.119& 0.259    &  1.024 &0.097 & 0.256 &  0.990 & 0.061& 0.269  \\
$\alpha_{23}= 1.232$& 1.288&0.045&0.313 & 1.340  &0.087& 0.381   &  1.308 & 0.061& 0.259 &  1.325 & 0.075& 0.407   \\
$\alpha_{24}= 1.634$& 1.432&-0.124&0.374 & 1.781 &0.090& 0.573   &  1.475 & -0.097& 0.362 &   1.981 & 0.213& 0.924  \\
$\psi_{12}= 0.469$&   0.471& 0.002 &0.132 &0.500 &0.065&0.105   &  0.465 & -0.009 & 0.150 &  0.479  & 0.021  &  0.111    \\
\hline
\emph{Avlog-lik} &\multicolumn{3}{c}{-1216.80}&\multicolumn{3}{c}{-1213.50}&\multicolumn{3}{c}{-1215.69}&\multicolumn{3}{c}{-1213.16}\\
\emph{\%cv}&\multicolumn{3}{c}{99}&\multicolumn{3}{c}{100}&\multicolumn{3}{c}{99}&\multicolumn{3}{c}{100}\\
\emph{Nrfeval} &\multicolumn{3}{c}{34.13}&\multicolumn{3}{c}{34.61}&\multicolumn{3}{c}{37.75}&\multicolumn{3}{c}{42.51}	\\
\emph{Avtime (s)} &\multicolumn{3}{c}{92.55}&\multicolumn{3}{c}{121.44}&\multicolumn{3}{c}{124.38}&\multicolumn{3}{c}{192.32}\\\hline
\end{tabular}
\end{table}

\begin{figure}
\psfrag{a11}{\tiny$\alpha_{11}$}\psfrag{a12}{\tiny$\alpha_{12}$}\psfrag{a21}{\tiny$\alpha_{21}$}\psfrag{a22}{\tiny$\alpha_{22}$}\psfrag{a23}{\tiny$\alpha_{23}$}\psfrag{a24}{\tiny$\alpha_{24}$}\psfrag{a35}{\tiny$\alpha_{35}$}\psfrag{a36}{\tiny$\alpha_{36}$}
\psfrag{a47}{\tiny$\alpha_{47}$}\psfrag{a48}{\tiny$\alpha_{48}$}
\psfrag{ps12}{\tiny$\psi_{12}$}\psfrag{ps13}{\tiny$\psi_{13}$}\psfrag{ps14}{\tiny$\psi_{14}$}\psfrag{ps23}{\tiny$\psi_{23}$}\psfrag{ps24}{\tiny$\psi_{24}$}\psfrag{ps34}{\tiny$\psi_{34}$}\psfrag{s}{\tiny$s$}
\makebox{\includegraphics[height=4cm,width=4cm]{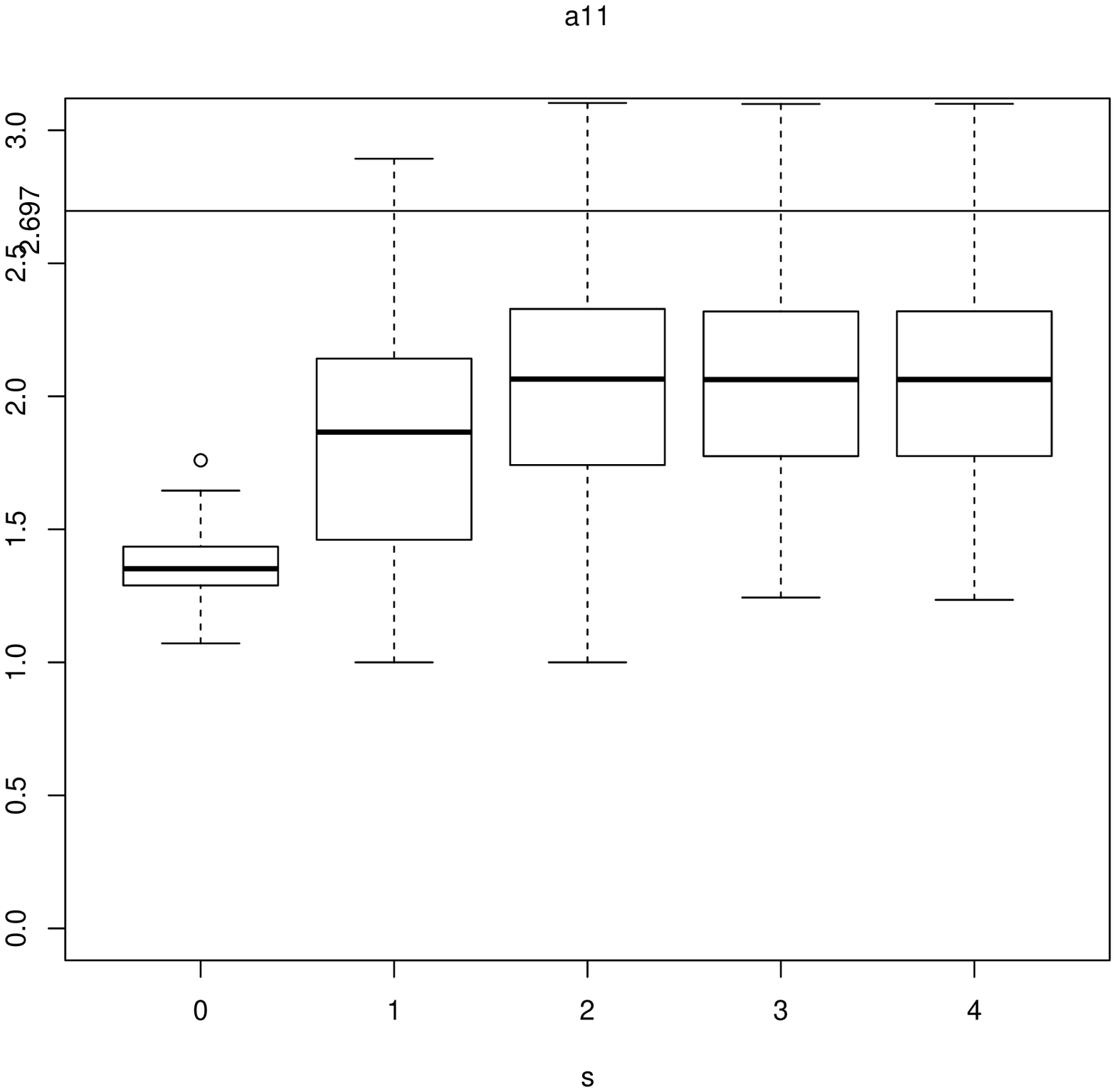}
\includegraphics[height=4cm,width=4cm]{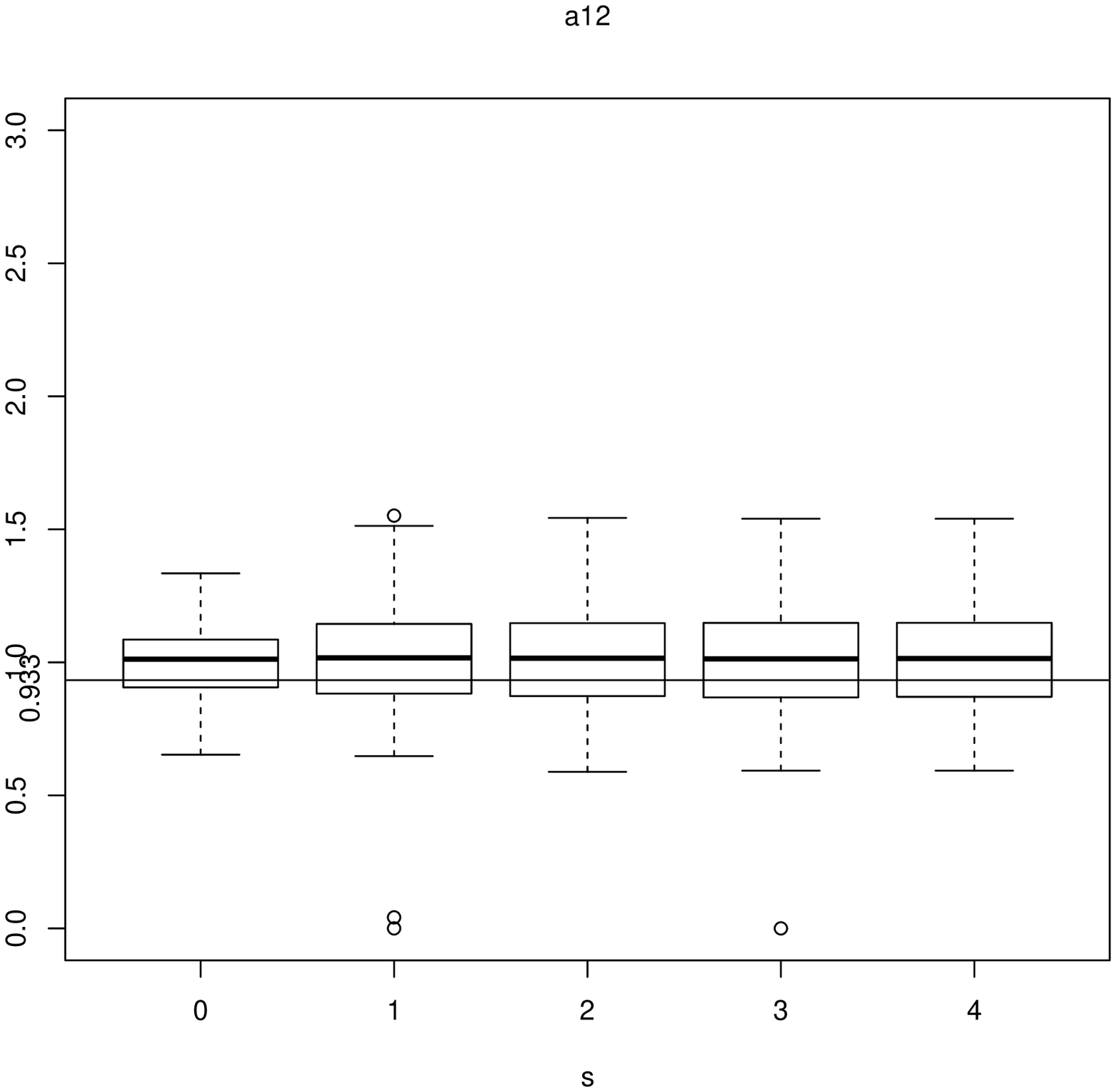}\includegraphics[height=4cm,width=4cm]{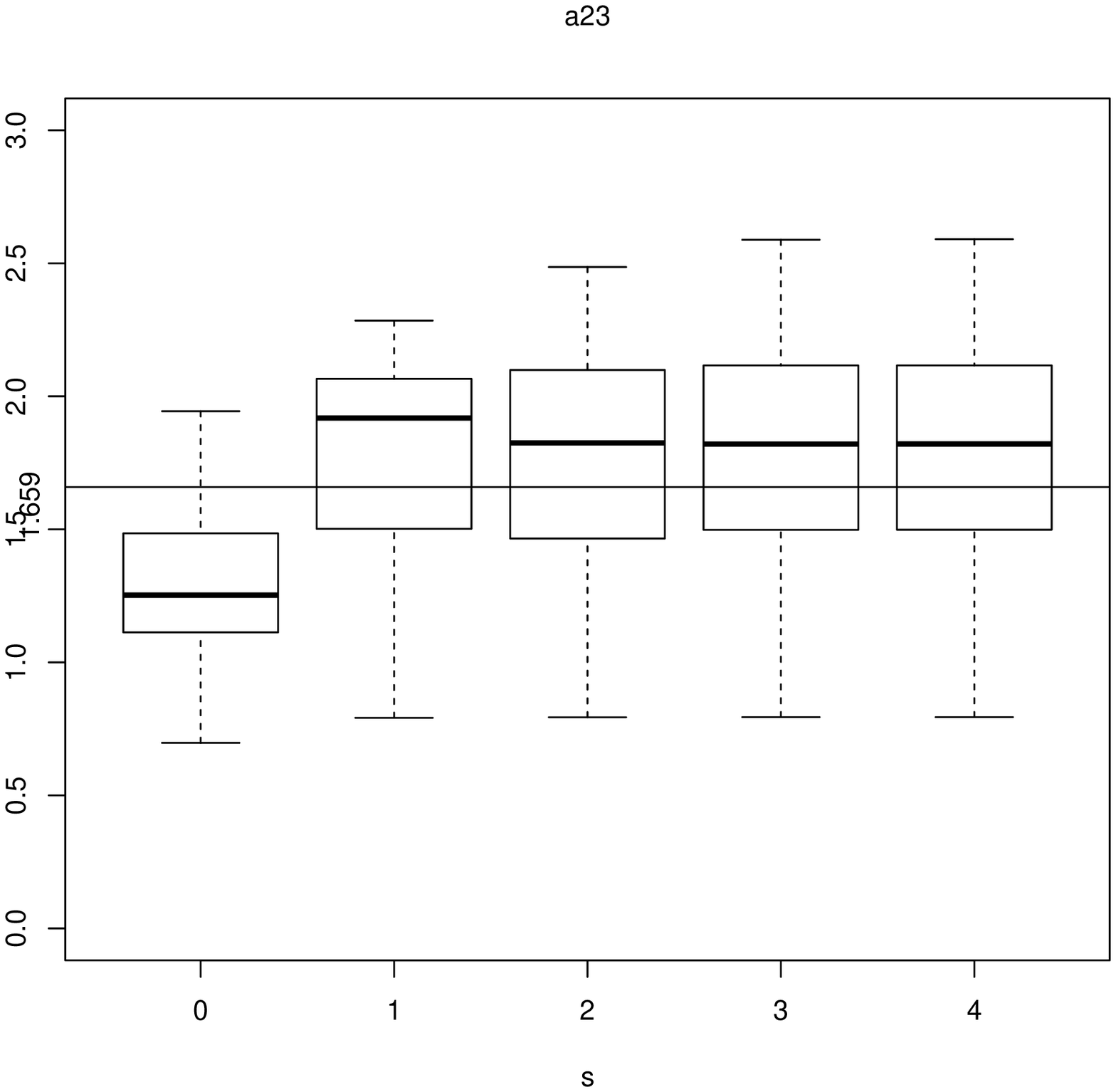}\includegraphics[height=4cm,width=4cm]{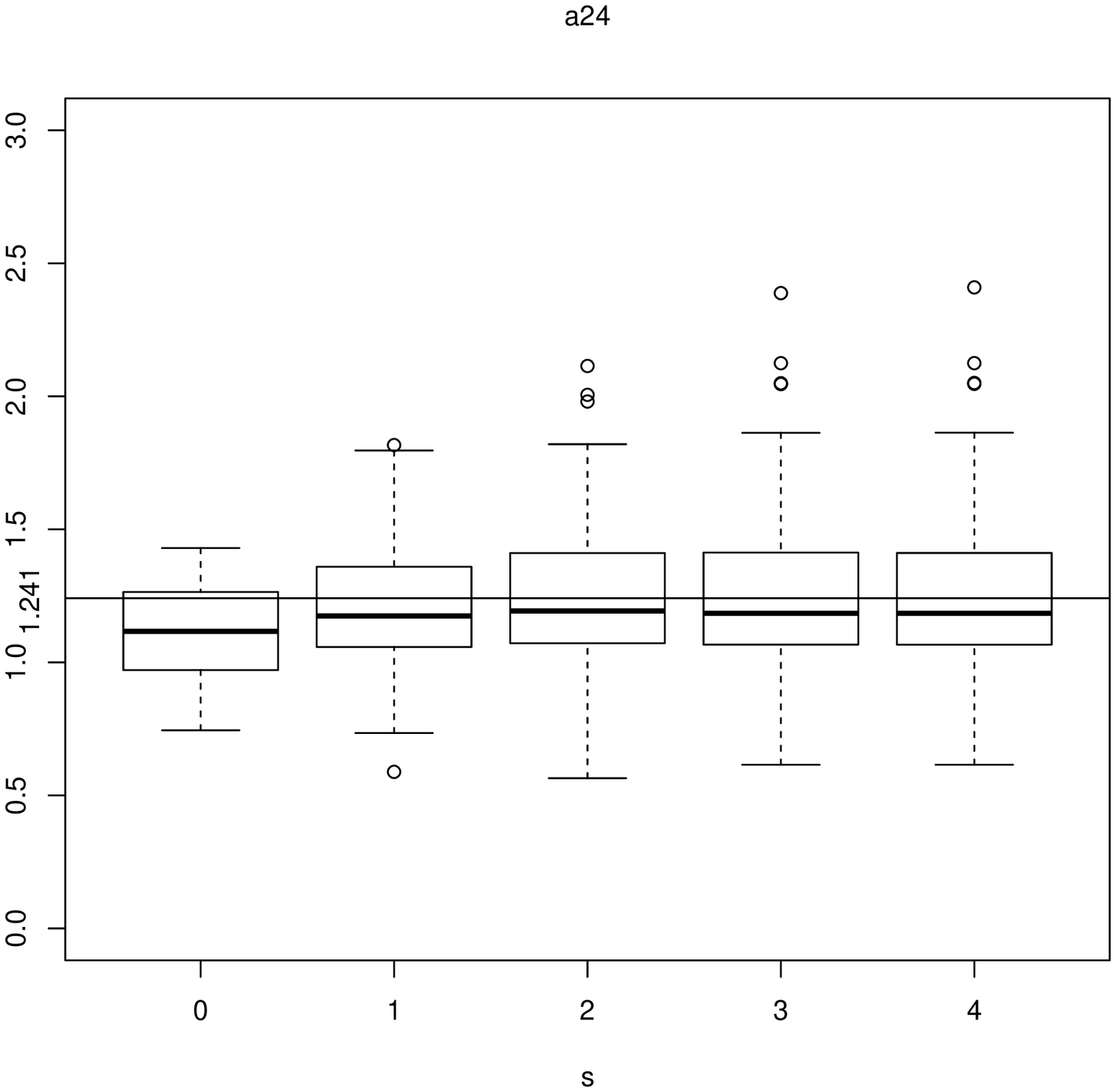}}

\makebox{\includegraphics[height=4cm,width=4cm]{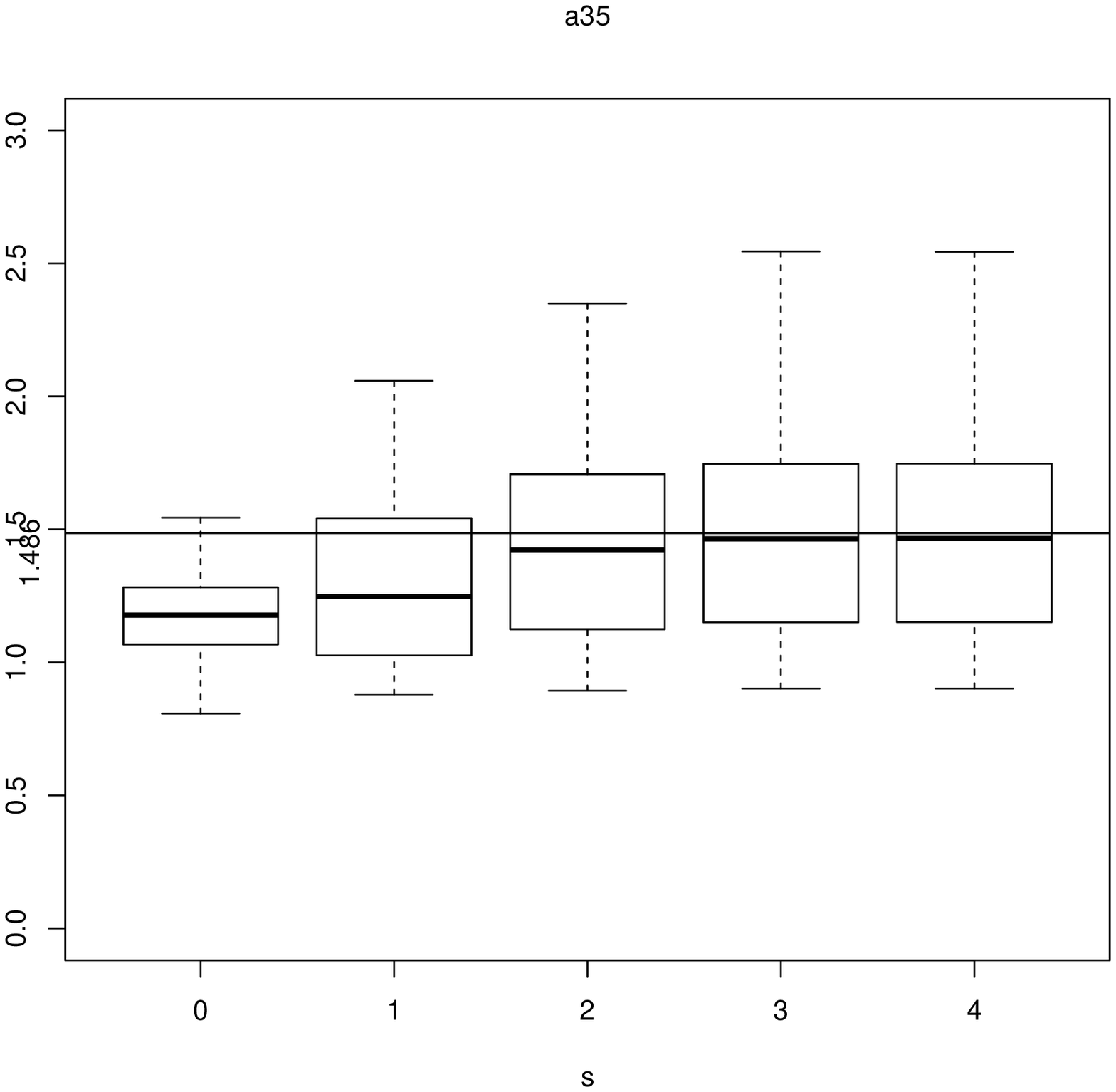}
\includegraphics[height=4cm,width=4cm]{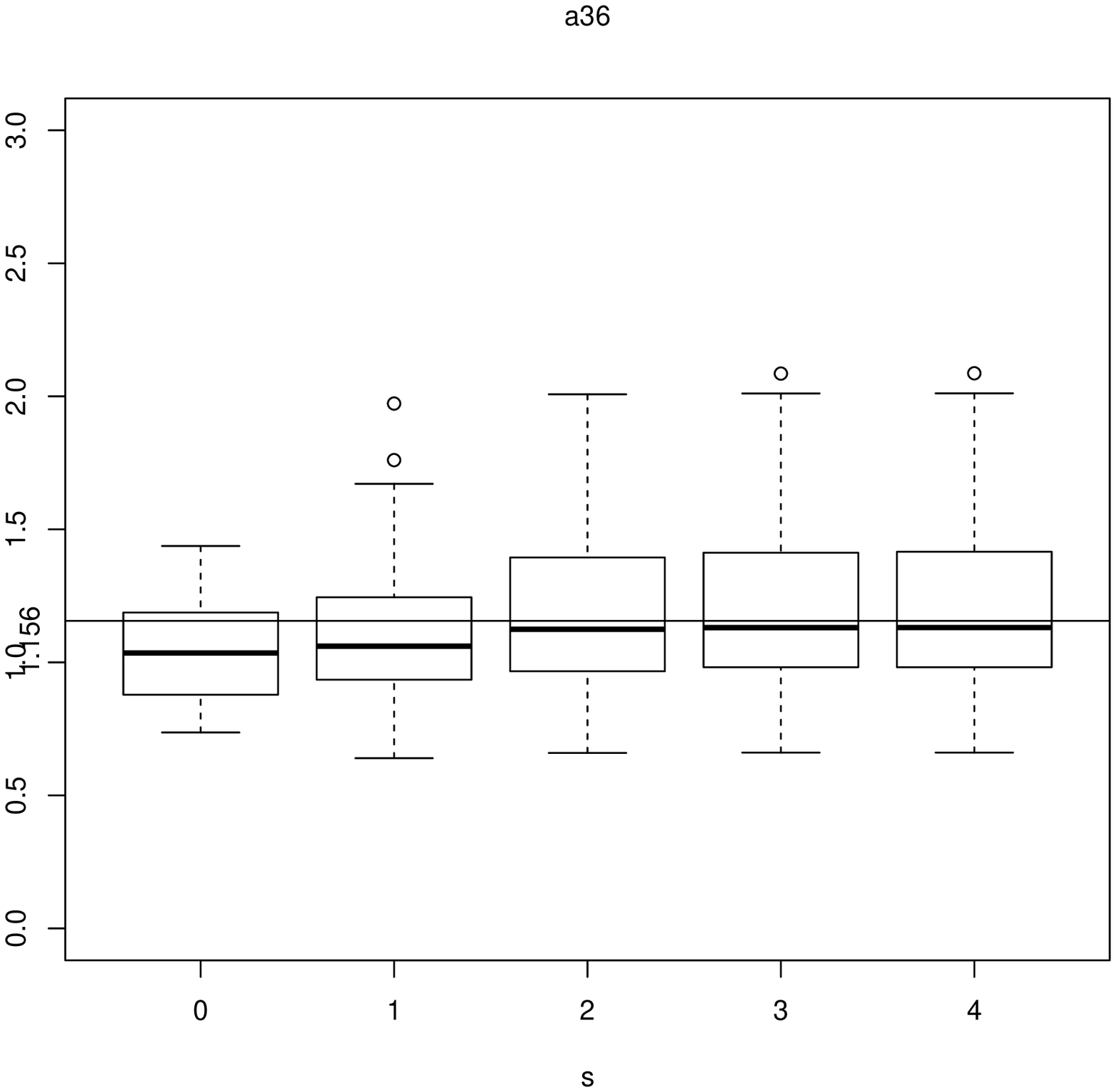}\includegraphics[height=4cm,width=4cm]{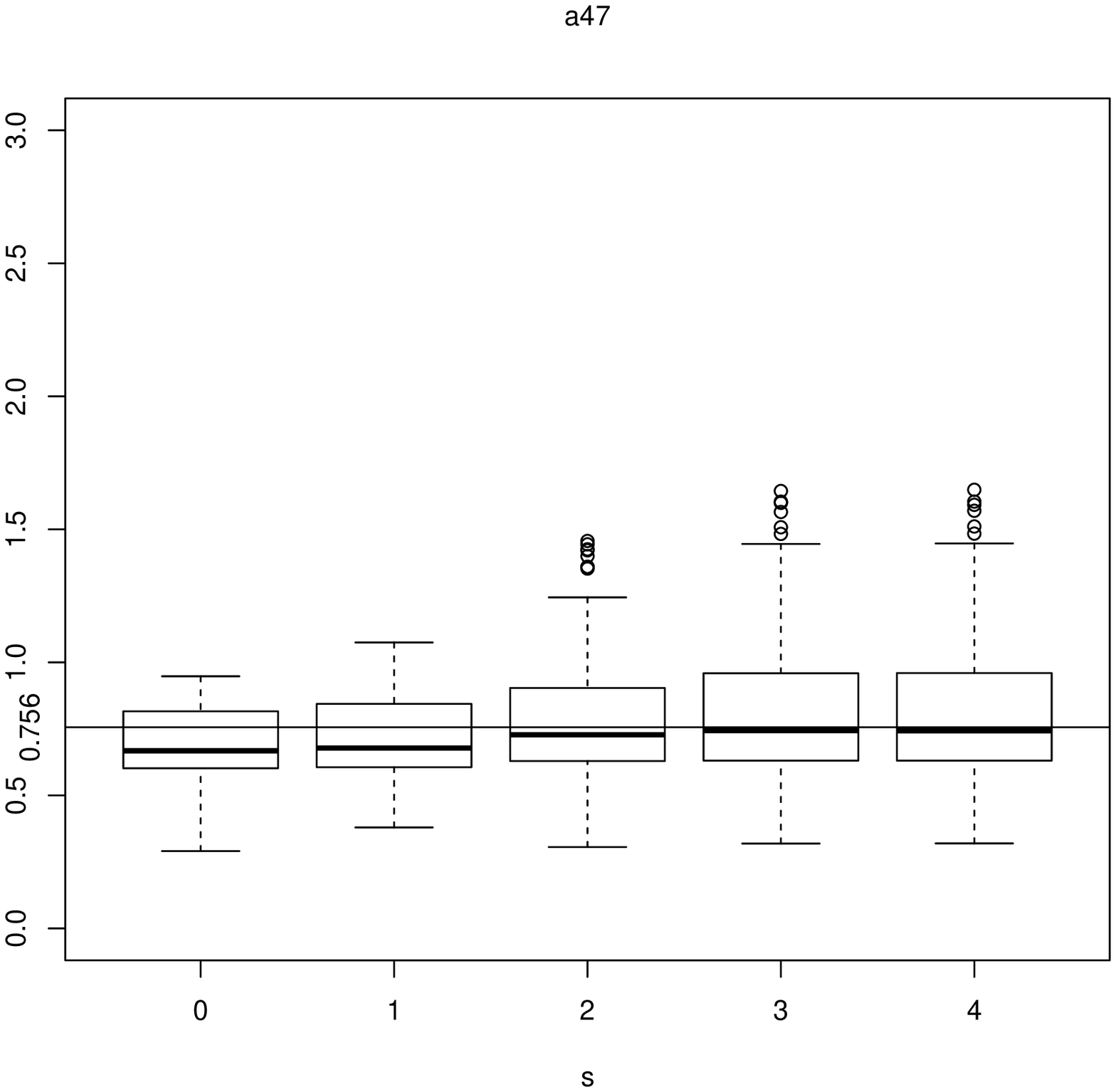}\includegraphics[height=4cm,width=4cm]{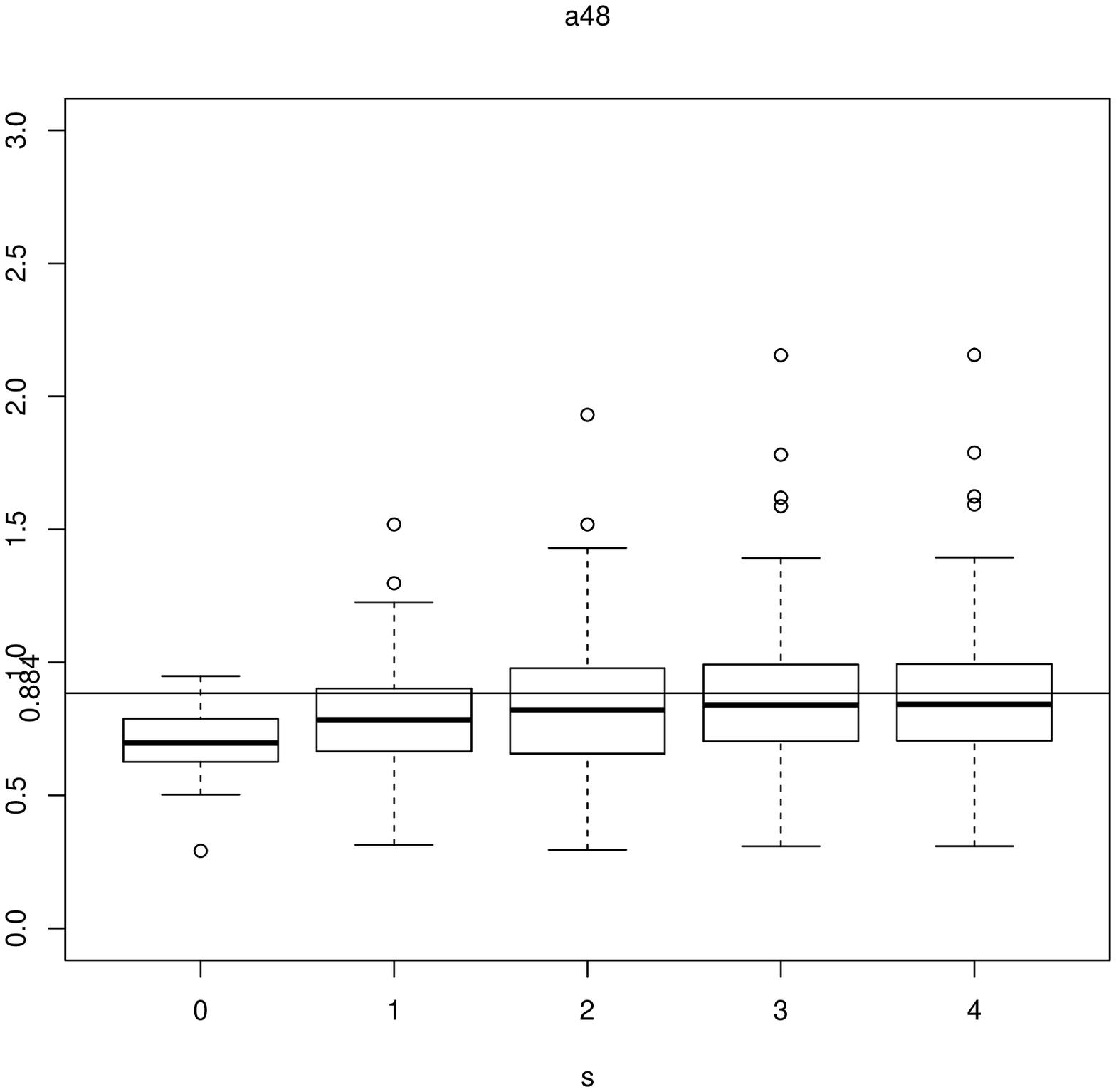}}

\makebox{\includegraphics[height=4cm,width=4cm]{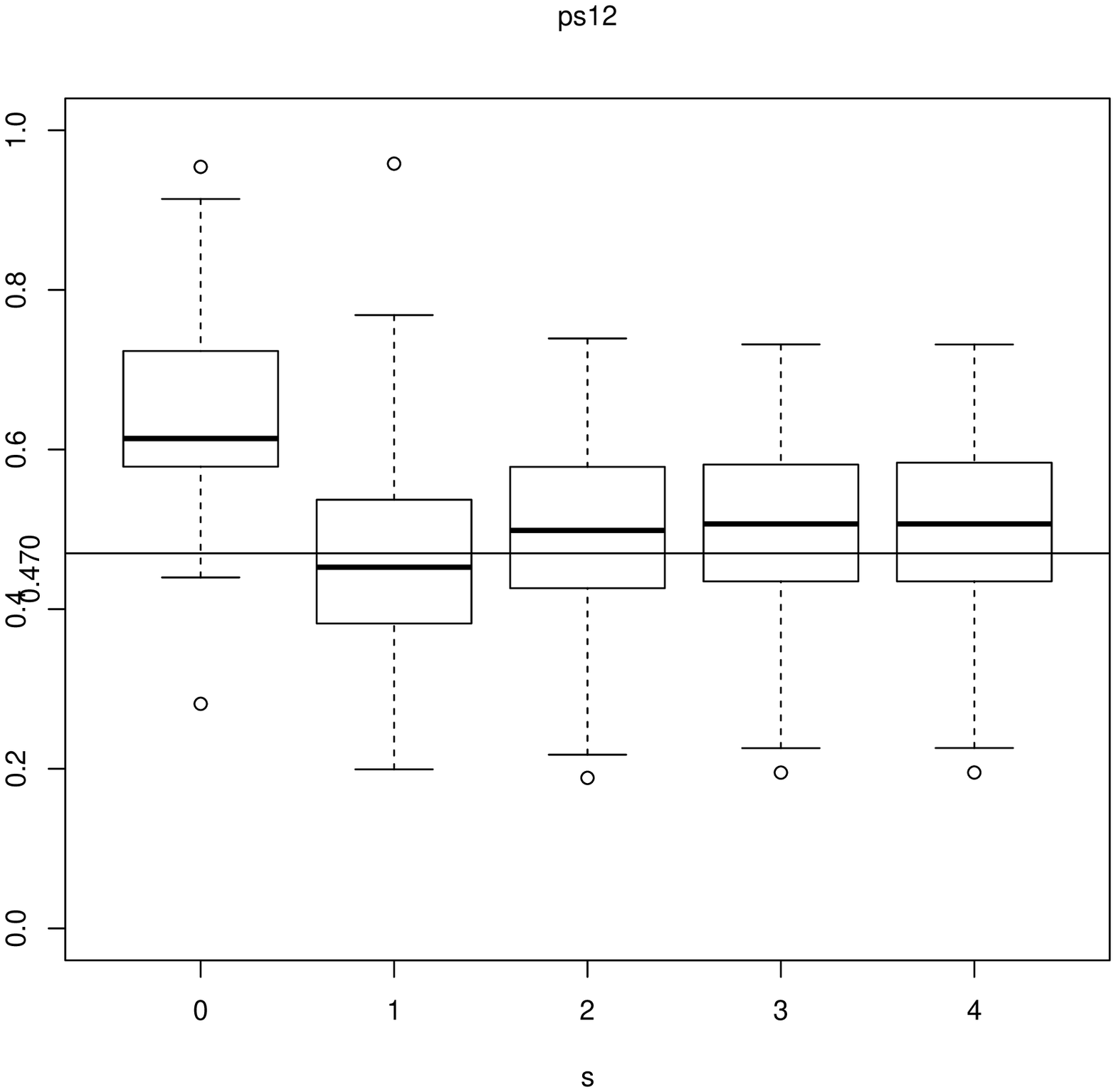}\includegraphics[height=4cm,width=4cm]{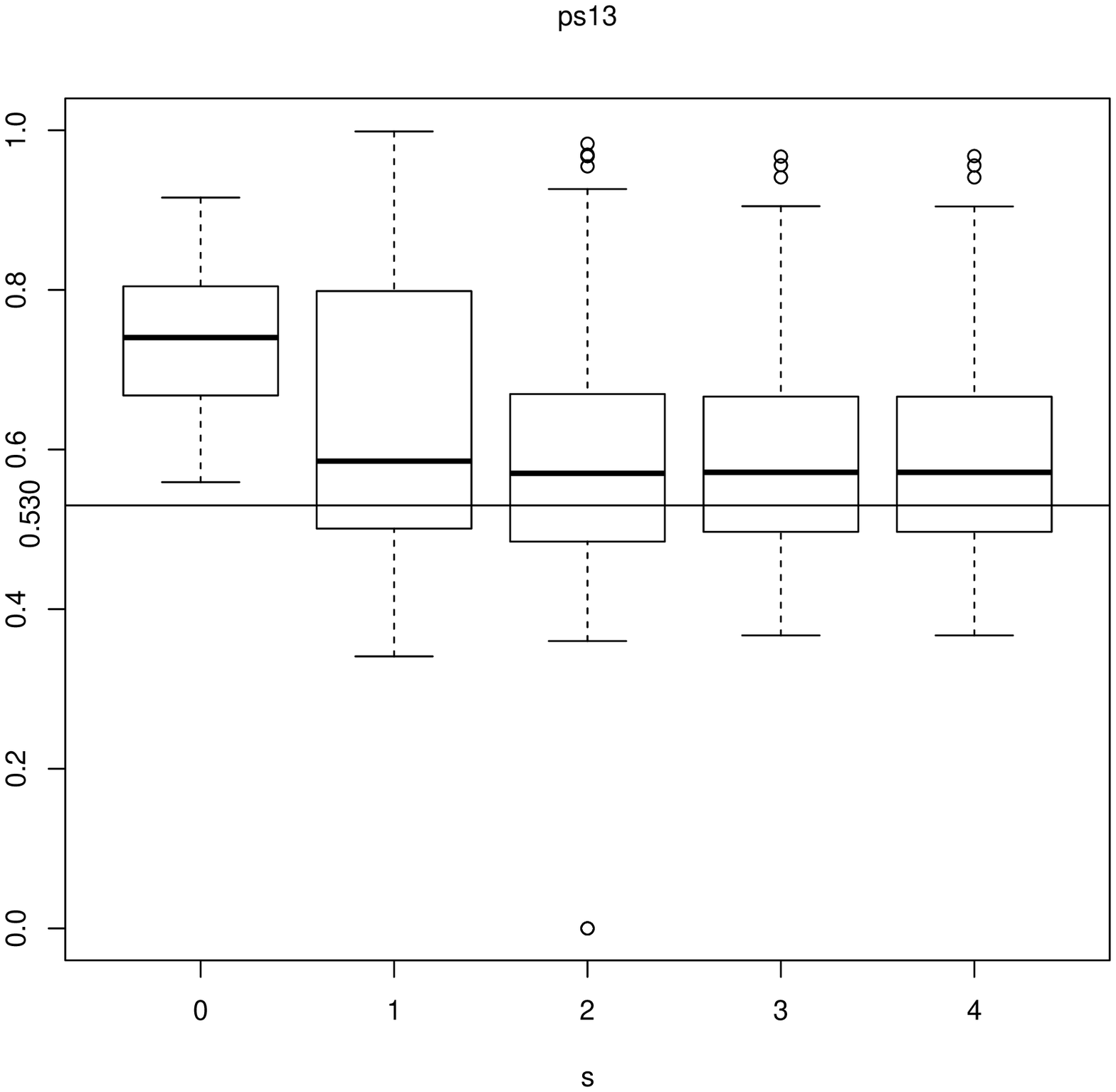}\includegraphics[height=4cm,width=4cm]{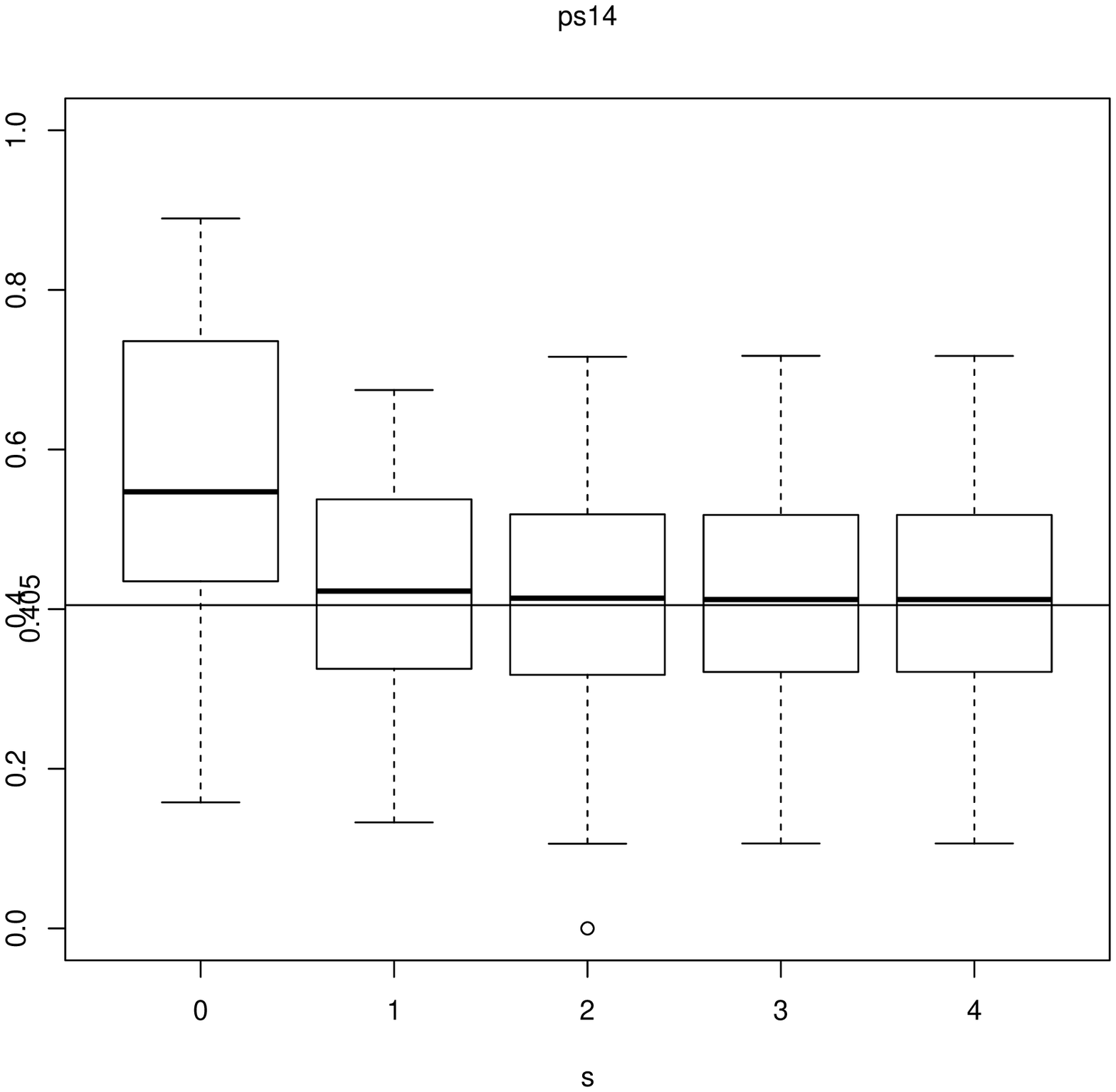}\includegraphics[height=4cm,width=4cm]{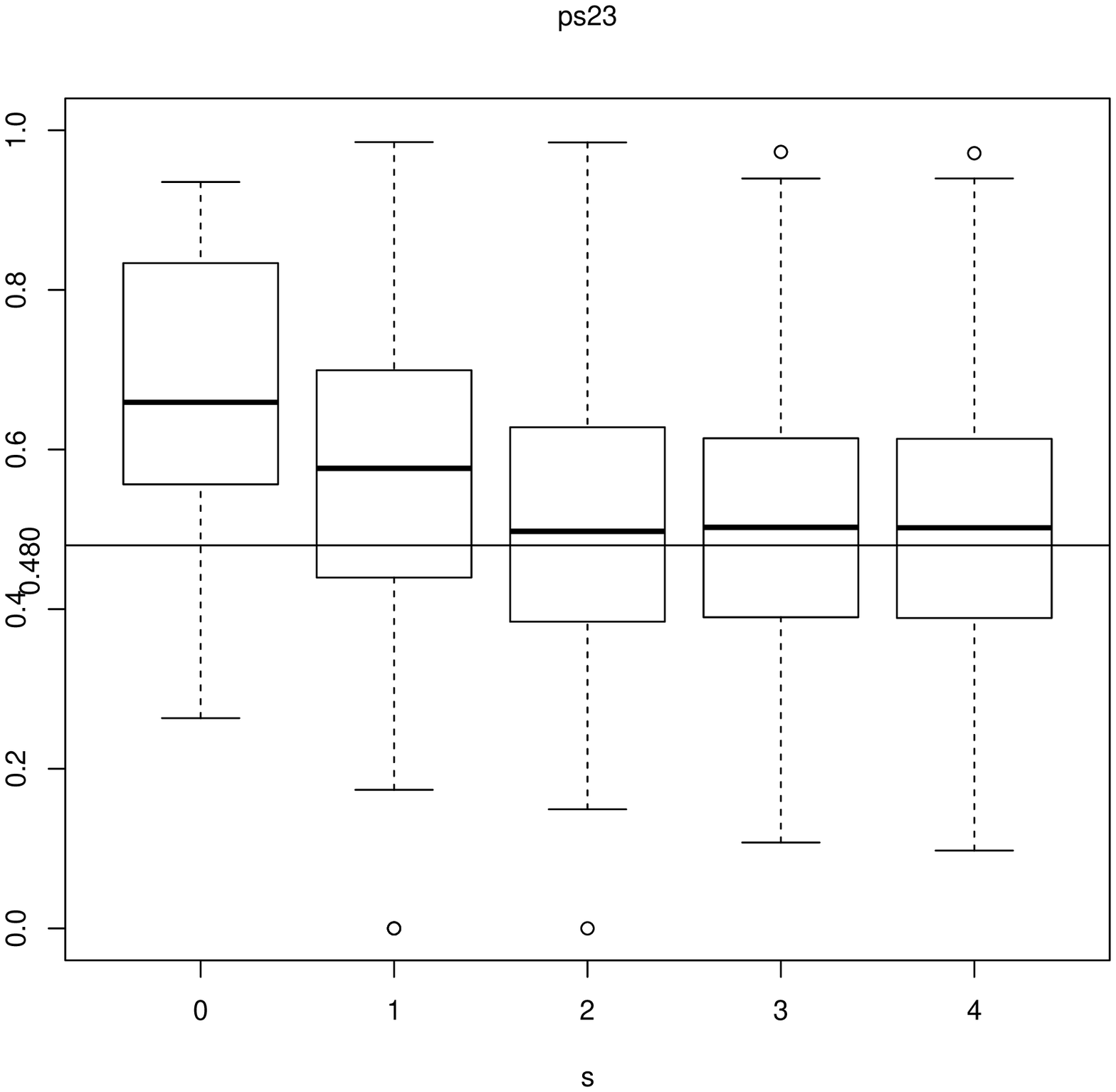}} 

\makebox{\includegraphics[height=4cm,width=4cm]{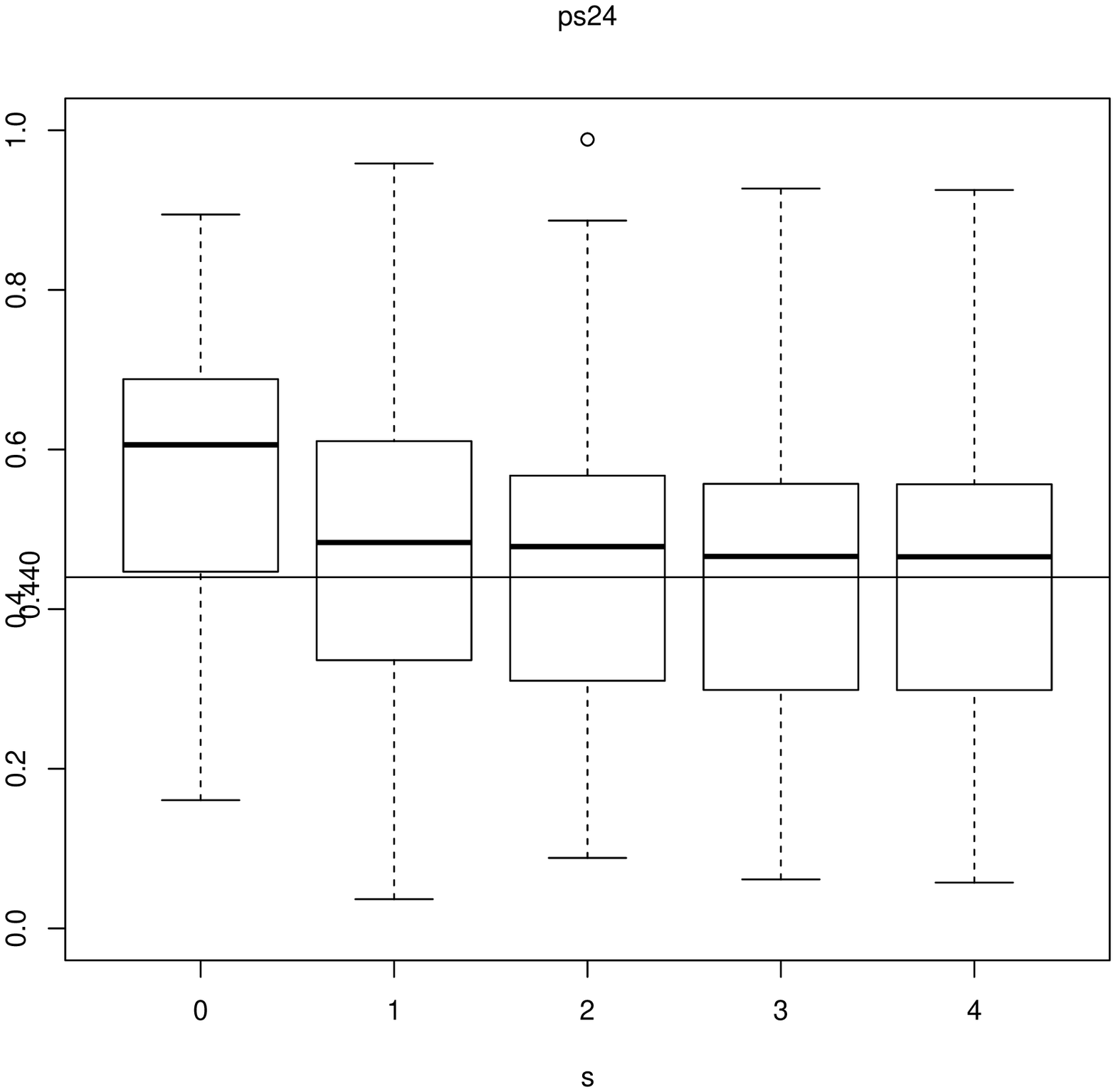}\includegraphics[height=4cm,width=4cm]{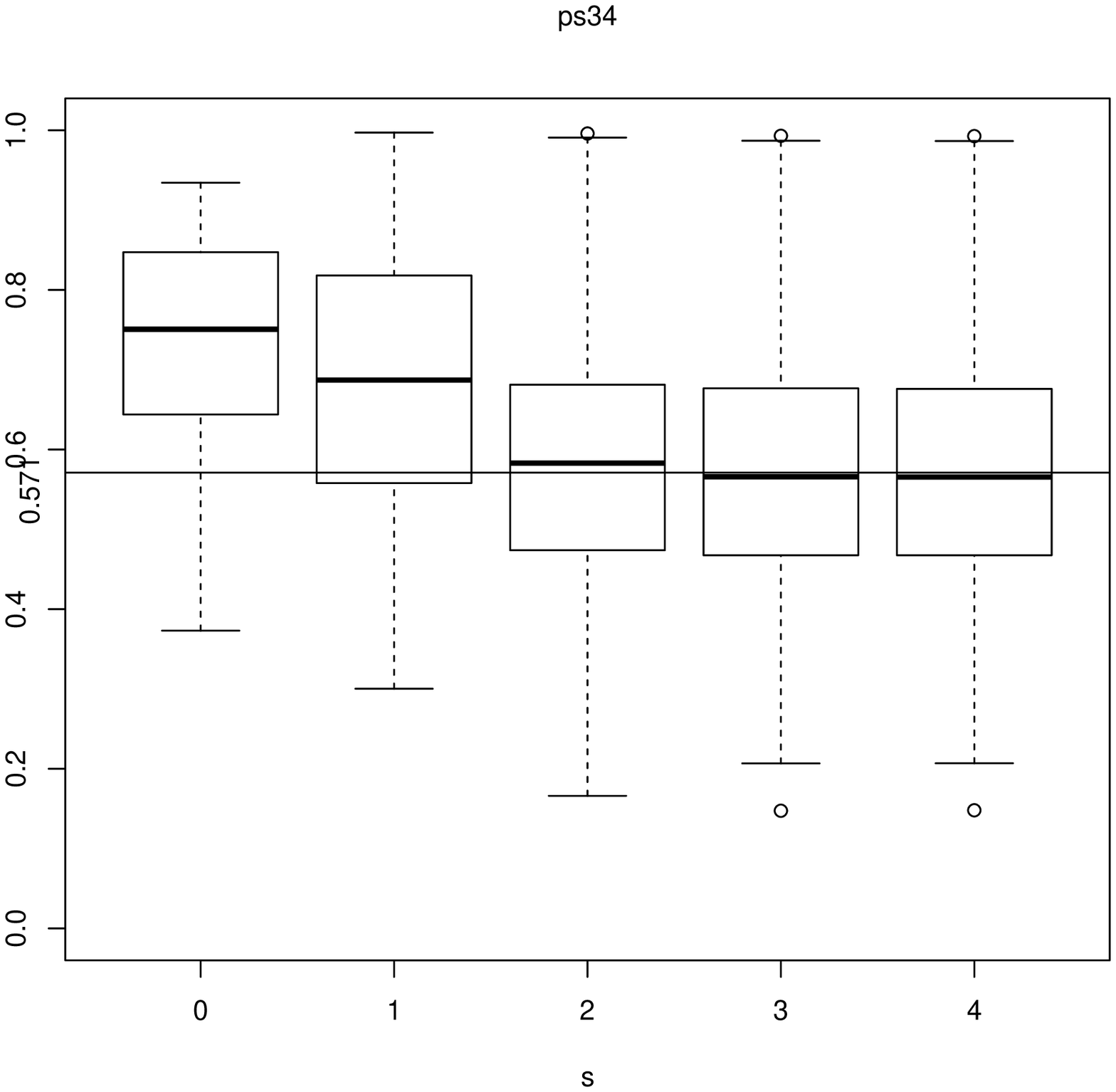}}
\caption{Box-plots of the parameter estimates for the four-factor model under Laplace ($s=0$), DRM (for $s=1,2$, and 3), and AGH ($s=4$), for $n_{q}=5, n=500$, 100 replications.}\label{fig02}
\end{figure}

In the second scenario, we evaluate the performance of DRM for all possible values of $s$, that is zero, one, two, three. This allows to compare, also in this case, the
results of DRM with Laplace ($s=0$) and AGH. The latter is feasible only if we set the number of nodes for each dimension equal to five or seven. We consider the solution for $n_q=5$
since there are no relevant differences with $n_q=7$. Also in
this scenario, we consider a simple structure for the matrix of the loadings that is characterized by two free parameters per each factor. In particular, the loading parameters are fixed to
$\alpha_{11}=2.697, \alpha_{12}=0.933, \alpha_{23}=1.659, \alpha_{24=}1.241, \alpha_{35}=1.486, \alpha_{36}=1.156, \alpha_{47}=0.756,\alpha_{48}=0.884$  and the correlation parameters are fixed
to $\psi_{12}=0.470,\psi_{13}=0.534, \psi_{14}=0.480,\psi_{23}=0.405,\psi_{24}=0.440,\psi_{34}=0.571.$
In Figure \ref{fig02} the box-plots of the Monte Carlo estimates are illustrated.

\begin{table}\footnotesize
\caption{\label{tab_comp} Computational performance of the algorithms based on the Laplace approximation, DRM and AGH  based on five quadrature points for the four-factor model with $n=500$,100 replications}
\centering%
\begin{tabular}{lrrrrrrrrrrrrrrr} \hline
&&&&&&&&&&&&\\
&\multicolumn{3}{c}{\emph{Laplace}} &\multicolumn{9}{c}{\emph{DRM5}}&\multicolumn{3}{c}{\emph{AGH5}} \\\hline
&\multicolumn{3}{c}{\emph{s=0}} &\multicolumn{3}{c}{\emph{s=1}}&\multicolumn{3}{c}{\emph{s=2}}&\multicolumn{3}{c}{\emph{s=3}} &\multicolumn{3}{c}{}\\ \hline
\emph{Avlog-lik} &\multicolumn{3}{c}{-2404.76}&\multicolumn{3}{c}{-2398.89}&\multicolumn{3}{c}{-2394.77}&\multicolumn{3}{c}{-2394.16}&\multicolumn{3}{c}{-2393.83}\\
\emph{\%cv}&\multicolumn{3}{c}{26}&\multicolumn{3}{c}{84}&\multicolumn{3}{c}{98}&\multicolumn{3}{c}{100}&\multicolumn{3}{c}{100}\\
\emph{Nrfeval}& \multicolumn{3}{c}{26.27}& \multicolumn{3}{c}{26.79}&\multicolumn{3}{c}{28.32}& \multicolumn{3}{c}{30.36}& \multicolumn{3}{c}{36.98}\\
\emph{Avtime (s)}& \multicolumn{3}{c}{240.75}& \multicolumn{3}{c}{290.12}&\multicolumn{3}{c}{438.43}& \multicolumn{3}{c}{902.06}& \multicolumn{3}{c}{909.13}\\ \hline
\end{tabular}
\end{table}

The accuracy of the parameter estimates is in line with the findings of the previous scenario. The  bias of the estimates reduces as $s$ increases, whereas the variability of the loadings is underestimated.  In terms of correlation estimates, the bias decreases as $s$ increases, with a similar behavior for the variability of the estimates when $s$ is equal to two and three and for the AGH case. It is  worth noting that, in this model, the  bias and variability of the estimates obtained with DRM5 with $s=2$ and with $s=3$ are very close, and they are coincident for DRM5 with $s=3$ and  AGH5 for the most of the parameter estimates. Hence, the information that we gain passing from $s=2$ to $s=3$ is small and passing from $s=3$ to the AGH approximation  is almost irrelevant.
The computational performance of the algorithm based on the Laplace approximation, DRM with $s=2$ and  3, and AGH  is analyzed in Table \ref{tab_comp}. We can observe that the Laplace method
performs very poorly in terms of samples in which the algorithm converges properly (only 26\%). This percentage rises to 84\% under DRM5 with $s=1$, to 98\% under DRM5 with $s=2$ and it is 100\%
under DRM5 with $s=3$ and under AGH5.   But if we look at the computational time
 we can observe that, on average, when $s=2$, the algorithm requires half of the time to converge compared with $s=3$ and AGH5 that are very similar also in terms of computational time.

\section{HRS data analysis}

Every two years the Institute for Social Research at the University of Michigan conducts a national representative Health and Retirement Study (HRS) funded by the National Institute on Aging.
The study is designed with the aim to investigate
the health, social and economic implications of the aging of the American population (Juster and Suzman, 1995). Between August and December 2001, a random sample of 856 subjects from all regions
of the country aged 70 or older was selected from the total sample  of approximately 22000 HRS individuals in order to participate to a more extensive study on Aging, Demographics And Memory
Study (ADAMS). The latter represents the first population-based study of dementia in the United
States (the data are available on the website http://www.rand.org). The participants received a thorough in-home clinical and neuropsychological assessment that allowed researchers to estimate
the prevalence, predictors, and outcomes of dementia in the U.S. elderly population. Three waves of data were collected from 2001 to 2008. \\
The aim of our analysis is to evaluate the performance of the proposed DRM approach  by first fitting a GLLVM for binary data to the ADAMS data available at the first time point and then by
conducting a longitudinal analysis over the available time points of one of the  dimensions detected in the first analysis.
\subsection{Factor analysis of the HRS data set}
The individuals selected for the ADAMS study were asked to answer to the items of the Section D of the HRS questionnaire devoted to the cognitive assessment. A detailed description of the
measures of the cognitive functioning can be found in \cite{OfFiHe}. For our analysis, we selected a subset of these measures corresponding to the following binary items:

\begin{enumerate}
  \item [1.] Please tell me today's date: year [MSE1]
  \item [2.] Please tell me today's date: date [MSE2]
  \item [3.] Please tell me today's date: day of week  [MSE3]
  \item [4.] Please tell me today's date: month   [MSE4]
  \item [5.] Let's try some subtraction of number: 100 minus 7 equals what? [SER71]
  \item [6.] And 7 from that? [SER72]
  \item [7.] And 7 from that? [SER73]
  \item [8.] And 7 from that? [SER74]
  \item [9.] And 7 from that? [SER75]
  \item [10.] Who is the President of the United States right now? [PRE]
  \item [11.] Who is the vice President? [VPRE]
  \item [12.] Please try to count backward as quickly as you can from 20 to 11 without error [BW20]
  \item [13.]  Please try to count backward as quickly as you can from 86 to 77 without error [BW86]
\end{enumerate}

For all of them, 1 means \textit{correct answer} and 0 means \textit{wrong answer}. Previous exploratory and confirmatory analysis on the HRS data \citep{McArFiKa,McAr} highlighted that these
items are measure of four latent factors (sub-scales) that are expression of the mental status of the interviewed people. In more detail, items from 1 to 4 are measures of the sub-scale Dates
(DA), items 10 and 11 are measures of the sub-scale Names (NA), items 12 and 13 are measures of the sub-scale Backward Counting (BC) and items from 6 to 9 are measures of the sub-scale Serial 7's
Test (S7).  Thus, we fitted a four-factor model as defined in (\ref{cond}), assuming a
simple structure for the matrix of loadings and, for the identifiability reasons discussed before, considering $\boldsymbol \Psi$ as correlation matrix. We allowed $s$ to vary from zero to three.
The solution for $s$ equal to 0 corresponds to the Laplace approximation and the solution when no DRM is performed, that  is the integral is four-dimensional, is obtained by applying the AGH. 
Using five and seven quadrature points, for both DRM and AGH, we obtained very similar results, thus Table \ref{tab03} shows the results only for $n_q=5$. 

Table \ref{tab03}  gives the estimated loading parameters with associated asymptotic standard errors obtained from (\ref{Mest}). 
The magnitude of the loading estimates is close for all values of $s$, whereas, in general, the smallest standard errors are obtained with $s=3$ and AGH. All the loadings have values
greater than one and are significantly different from zero independently on the approximation used. Thus, the items are all relevant measures of the corresponding underlying factors. 
Table \ref{tab04} reports  the estimated correlation parameters, the value of the log-likelihood and the computational time for the DRM with each value of $s$ and for the AGH. It is worth noting that the values of the
correlation parameters are very close and, in some cases, they are coincident at the second digit for all $s$.
 The smallest standard errors are obtained for $s=1$ and $s=3$, but, also in this case, we can observe that the significance of the estimates does not change with $s$. As expected, the
 log-likelihood increases with $s$, whereas the computational time for $s=3$ and AGH is more than twice than the one obtained with $s=2$.  Thus, we can conclude that $DRM5$ with $s=2$ seems to provide  the best compromise in terms of accuracy of the estimates and computational performance.
\subsection{Longitudinal analysis of the HRS data set}
Multidimensional longitudinal data can be challenging to analyze since a large number of random effects/latent variables can be required to take into account for different sources of
variability present in the data. In this section, we analyze the items  $SER71$ $SER72$ $SER73$ $SER74$ $SER75$ of the ADAMS data set observed at three different time points (2001-2003,
2002-2005, 2006-2008). The aim is to
investigate how the working memory and mental processing task in which
respondents counted backward from 100 by 7s changes over time.\par
We applied the GLLVM for multidimensional longitudinal data proposed by \cite{Dun03} and \cite{CaMouVa09} to analyze $p$ binary variables  observed at $T$ different time points, such that
$\mathbf{y}_{l}=(y_{1t},y_{2t},\ldots,y_{pt})$  is a $pT$- dimensional vector. Beyond the association  between several items at the same time point, their dependence over time as well as
the variability of the same item at different time points have to be considered. The latter is accounted by $p$ item-specific random effects $\mathbf{u}_{l}=(u_{1l}, \ldots, u_{pl})^{T}$, whereas
the
former is explained by means of $T$ time-dependent latent factors $\mathbf{z}_{l}=(z_{1l}, \ldots , z_{Tl})$,  such that the vector of latent variables $\mathbf{b}_{l}=(\mathbf{z}_{l},
\mathbf{u}_{l})^{T}$ is $(p+T)$-dimensional. The correlation of the latent variables $\mathbf{z}_{l}$ over time is modeled through a first order non-stationary autoregressive process of parameter
$\phi$
and
$Var(z_1)=\sigma^2_1$. Hence, the covariance matrix of $\mathbf{b}_{l}$, $\boldsymbol \Psi$, is a block matrix $\left[
\begin{array}{cc}
\boldsymbol \Gamma & 0 \\ 0 & \boldsymbol {\Sigma_u}
\end{array}
\right] \label{cv1}$
with $\boldsymbol{\Sigma_u}=diag_{i=1,\ldots,p}\{\sigma_{u_{i}}^2\}$.
 The elements of $\boldsymbol \Gamma$ are the variances and covariances of the time-dependent latent variables $z_t$, that is
$\gamma_{t,t}=Var(z_t)=\phi^{2(t-1)}\sigma_1^2+I(t\geq 2)\sum_{k=1}^{t-1}{\phi^{2(k-1)}}$ and
$\gamma_{t,t'}=Cov(z_t,z_{t'})=\phi^{t+t'-2}\sigma_1^2+I(t\geq 2)\sum_{k=0}^{t-2}{\phi^{t'-t+2k}}$, where $I(.)$ is the indicator function and $t<t'$.\\The measurement part of the model has the same specification given in eq. (\ref{cond}), but the linear predictor is now characterized by a $pT$-dimensional vector $\boldsymbol \alpha_{0}$ of
item- and time-specific intercepts, and a loading matrix $\boldsymbol \alpha$ of dimension ($pT \times (p+T)$) whose generic submatrix referring to the $j$-th item is given by
$\left[\alpha_{j}\mathbf{I}_{T} \quad \mathbf{O}[j, \mathbf{1}_{T}]\right]$, being  $\mathbf{I}_{T}$  the identity matrix of dimension $T \times T$, and $\mathbf{O}[j, \mathbf{1}_{T}]$ a null
matrix of dimension $T \times (T+1)$ whose $j$-th column is substituted by a $T$-dimensional vector of ones.\\
It can be noticed that, in the longitudinal setting, the number of latent variables increases linearly with the number of observations. In this particular example, the number of latent variables
and random effects is equal to eight ($q=8$), being $p=5$ and $T=3$. It follows that AGH is unfeasible also using five quadrature points per each dimension, hence only DRM can be applied in this case. We
consider a fully constrained model (equal thresholds and equal loadings over time) in order to obtain the same metric (origin and scale) of the latent variable over the three time points. The
resulting strict measurement invariance allows comparisons of the factor across time.  

As for the choice of $s$, we started fitting the model with $s=1$ and we increased its value until the
mean
of the absolute differences between parameter estimates ($\emph{Av}(\triangle$)) became sufficiently small (order of $10^{-3}$).\\ 
In Table \ref{tab05} we report the estimated model parameters under DRM with five quadrature points for $s=1,2,3,4$.
It is interesting to notice that, for this particular model, apart from $\phi$, all the parameter estimates change sensibly from $s=1$ to $s=2$ ($\emph{Av}(\triangle_{12}$)=0.164). On the
contrary, they are all very close for $s=3$ and $s=4$ ($\emph{Av}(\triangle_{34}$)=0.007), indicating that the additional information included in the likelihood for $s>3$ is almost irrelevant.
From a computational point of view, there is a substantial difference
in the time to convergence for increasing values of $s$. For example, the computational time for $s=4$ is more than six times the computational time for $s=3$. It is evident that in this case
$s=3$ can be considered as the reference solution. \\ Looking at the results obtained when $s$ equals 3 (Table \ref{tab06}), we can notice that the variances of the random effects are all
quite small and not significantly different from zero, indicating that there is no relevant heterogeneity of individuals on each item over time. On the other hand, the autoregressive parameter is very
high and significant ($\widehat{\phi}=0.99$ with standard error equal to 0.15), indicating a highly persistent latent process over time.

 \begin{landscape}
 \begin{table}\tiny
\caption{\label{tab03} Estimated factor loadings with asymptotic standard errors in brackets for the fourth factor model, s=0,1,2,3,4, ADAMS data set}
\begin{tabular}{lrrrrrrrrrrrrrrrrrrrrr} \hline
&&&&&&&&&&&&&\\
&\multicolumn{4}{c}{\emph{Laplace}} &\multicolumn{12}{c}{\emph{DRM5}}&\multicolumn{4}{c}{\emph{AGH5}} \\\hline
&\multicolumn{4}{c}{\emph{s=0}} &\multicolumn{4}{c}{\emph{s=1}}&\multicolumn{4}{c}{\emph{s=2}}&\multicolumn{4}{c}{\emph{s=3}} &\multicolumn{4}{c}{}\\
\hline
& $DA$ & $S7$& $NA$ & $BC$& $DA$ & $S7$& $NA$ & $BC$& $DA$ & $S7$& $NA$ & $BC$& $DA$ & $S7$& $NA$ & $BC$& $DA$ & $S7$& $NA$ & $BC$&\\
\hline\\
$\widehat{\alpha}_{\mathrm{MSE1}}$ & 1.92  &-&-&-& 2.10 &-&-&-&  1.89  &-&-&-&     1.95 &-&-&-&  1.96 &-&-  &- \\
               & (0.43)&-&-&-&  (0.29)    &-&-&-& (0.31) &-&-&-&    (0.17)   &-&-&-& (0.34) &- &- &- \\
$\widehat{\alpha}_{\mathrm{MSE2}}$& 1.45  &-&-&-& 1.60 &-&-&-&  1.49   &-&-&-& 1.51   &-&-&-& 1.51  &- &- &-\\
               & (0.52)&-&-&-&  (0.42)    &-&-&-& (0.51)  &-&-&-& (0.24)  &-&-&-& (0.39) &- &- &-\\
$\widehat{\alpha}_{\mathrm{MSE3}}$& 1.84  &-&-&-& 2.03 &-&-&-&  1.81   &-&-&-& 1.86   &-&-&-&  1.87  &-  &-  &-\\
               & (0.51)&-&-&-& (0.27)     &-&-&-& (0.43)  &-&-&-& (0.14) &-&-&-& (0.22) &- &-  &-\\
$\widehat{\alpha}_{\mathrm{MSE4}}$& 1.42&-&-&-& 1.54  &-&-&-& 1.41     &-&-&-& 1.44   &-&-&-& 1.45 &- &- &-\\
               &  (0.45) &-&-&-& (0.26)   &-&-&-& (0.35)   &-&-&-&(0.16) &-&-&-& (0.17)   &- &- &- \\
$\widehat{\alpha}_{\mathrm{SER71}}$& -& 1.76 &-&-&-&  1.82 &-&-&-& 1.77  &-&-&-&1.77    &-&-&-& 1.78 &- &- \\
               &-&  (0.45) &-&-&-& (0.36)    &-&-&-& (0.40) &-&-&-&(0.26) &-&-&-&  (0.39) &- &-\\
$\widehat{\alpha}_{\mathrm{SER72}}$&- &1.60  &-&-&-&  1.69 &-&-&-& 1.73   &-&-&-& 1.74  &-&-&-& 1.74 &- &- \\
               &-&   (0.83)  &-&-&-& (0.65)  &-&-&-& (0.39) &-&-&-&(0.28)&-&-&-&  (0.24) &- &-\\
$\widehat{\alpha}_{\mathrm{SER73}}$& -&1.18 &-&-&-& 1.26 &-&-&-& 1.30    &-&-&-& 1.30    &-&-&-& 1.28 &- &-\\
               &-&    (0.64) &-&-&-&(0.36)  &-&-&-& (0.29)  &-&-&-& (0.18)&-&-&-&  (0.19) &- &- \\
$\widehat{\alpha}_{\mathrm{SER74}}$& -&2.23  &-&-&-&2.28&-&-&-& 2.29    &-&-&-& 2.30    &-&-&-& 2.31 &- &- \\
               &-&    (0.91)&-&-&-& (0.42) &-&-&-& (0.63) &-&-&-& (0.35)&-&-&-&  (0.40) &- &- \\
$\widehat{\alpha}_{\mathrm{SER75}}$& -&1.71 &-&-&-&1.74 &-&-&-& 1.84     &-&-&-& 1.84  &-&-&-& 1.84  &- &-\\
                 &-&     (0.76) &-&-&-& (0.48) &-&-&-& (0.61) &-&-&-& (0.30)&-&-&-&  (0.38) &- &-\\
$\widehat{\alpha}_{\mathrm{PRE}}$ &- &-&1.12 &-&-&-& 1.35 &-&-&-&  1.12  &-&-&-& 1.17   &-&-&-&1.18  &- \\
                &-&-&  (0.91) &-&-&-&(0.27) &-&-&-&  (0.41) &-&-&-& (0.20) &-&-&-&  (0.23) &-  \\
$\widehat{\alpha}_{\mathrm{VPRE}}$  &-&-&1.39  &-&-&-&1.38 &-&-&-&1.43  &-&-&-& 1.45  &-&-&-&  1.45 &- \\
                    &-&-& (1.99) &-&-&-&(0.26) &-&-&-& (0.44) &-&-&-& (0.27)&-&-&-&  (0.22) &-  \\
$\widehat{\alpha}_{\mathrm{BW20}}$&-&-&-&1.43  &-&-&-&1.57 &-&-&-& 1.41   &-&-&-& 1.46  &-&-&-& 1.47  \\
                 &-&-&-& (0.37) &-&-&-&(0.23) &-&-&-& (0.59) &-&-&-&  (0.23)&-&-&-&  (0.22)  \\
$\widehat{\alpha}_{\mathrm{BW86}}$&-&-&-&1.05 &-&-&-&1.07 &-&-&-& 1.13   &-&-&-& 1.19  &-&-&-& 1.19\\
               &-&-&-&(0.21) &-&-&-&(0.28) &-&-&-& (0.36) &-&-&-& (0.20) &-&-&-&(0.26) \\
\end{tabular}
\end{table}
\end{landscape}

 \begin{table}
\caption{\label{tab04} Estimated correlation parameters with standard errors in brackets for the fourth factor model, s=0,1,2,3,4, ADAMS data set}\tiny \singlespacing
\begin{tabular}{lrrrrrr} \hline
&&&&&\\
&{\emph{Laplace}} &\multicolumn{3}{c}{\emph{DRM5}}&{\emph{AGH5}} \\\hline
&{\emph{s=0}} &{\emph{s=1}}&{\emph{s=2}}&{\emph{s=3}} &\\ \hline
$\widehat{\psi}_{12}$& 0.64 & 0.64&0.64 &0.64 &0.64\\
           & (0.07) &(0.02)&(0.06)&(0.01)&(0.27)\\
$\widehat{\psi}_{13}$& 0.91 & 0.92&0.92 &0.92 &0.92\\
           & (0.49) &(0.01)& (0.01)&(0.01)&(0.04)\\
$\widehat{\psi}_{23}$&0.55  & 0.51&0.57 &0.55 &0.55\\
           & (0.16) &(0.01)& (0.03)&(0.02)&(0.10)\\
$\widehat{\psi}_{14}$& 0.90 & 0.89&0.89 &0.89 &0.89\\
           &(0.31) &(0.02)& (0.02)&(0.01)&(0.07)\\
$\widehat{\psi}_{24}$& 0.85 & 0.85&0.83 &0.81 &0.81\\
           &(0.04)&(0.02)& (0.08)&(0.03)&(0.09)\\
$\widehat{\psi}_{34}$&  0.76 &0.77&0.74 &0.74 &0.74\\
           &(0.04)&(0.03)& (0.04)& (0.03)& (0.08) \\\\
\emph{log-lik} &-1413.01&-1410.90 &-1410.99& -1409.60& -1409.54\\
\emph{time (s)} &217.83& 157.82 & 238.80 & 592.02 & 552.10\\
\end{tabular}
\end{table}

\begin{table}
\caption{\label{tab05} Estimated model parameters under DRM5 for the non-stationary model and absolute relative differences between consecutive values of s, ADAMS data set}\tiny \singlespacing
\begin{tabular}{lrrrrrrr} \hline
& $s=1$ & $s=2$ &  $s=3$ & $s=4$ &$\triangle_{12}$& $\triangle_{23}$ &$\triangle_{34}$\\  \hline
$\widehat{\alpha}_{1}$&1.000 &1.000 &1.000&1.000&-    &  -  & - \\
$\widehat{\alpha}_{2}$&0.495 &0.551 &0.585&0.585&0.114&0.061&0.001   \\
$\widehat{\alpha}_{3}$&0.455 &0.508 &0.531&0.531&0.116&0.045&0.001   \\
$\widehat{\alpha}_{4}$&0.533 &0.586 &0.619&0.619&0.100&0.057&0.001   \\
$\widehat{\alpha}_{5}$&0.612 &0.678 &0.720&0.720&0.108&0.062&0.001   \\
$\widehat{\phi}$&                   0.995 &0.993 &0.994&0.994&0.002&0.001&0.001   \\
$\widehat{\sigma}^2_1$&             7.053 &6.052 &5.715&5.721&0.142&0.056&0.001   \\
$\widehat{\sigma}^2_{u_1}$&         0.560 &0.444 &0.573&0.562&0.206&0.291&0.020   \\
$\widehat{\sigma}^2_{u_2}$&         0.974 &1.157 &1.370&1.360&0.187&0.185&0.008   \\
$\widehat{\sigma}^2_{u_3}$&         0.466 &0.589 &0.741&0.734&0.265&0.257&0.009   \\
$\widehat{\sigma}^2_{u_4}$&         0.404 &0.512 &0.699&0.688&0.268&0.364&0.016   \\
$\widehat{\sigma}^2_{u_5}$&         0.241 &0.313 &0.404&0.399&0.301&0.290&0.014  \\
\emph{Av}($\triangle$)&          -  & -    & -   &  -  &0.164&0.152&0.007  \\
\emph{log-lik} &-520.26&-518.90&-517.75&-517.79&-&-&-\\
\emph{time (s)} &937.51&	1290.06& 6697.60&44246.78&-&-&-\\
                 \hline
\end{tabular}
\end{table} 

\section{Conclusions}

 One of the main problems in the estimation of GLLVMs is that integrals involved in the likelihood do not have an analytical solution due to the presence of  latent variables.   A ``gold'' standard approach
is represented by the Adaptive Gauss-Hermite quadrature, that is known to provide quite accurate estimates. However, it becomes computational unfeasible in presence of a large number of factors. A typical case is represented by GLLVMs for longitudinal data where the number of latent variables/random effects increases proportionally with the number of items.

In this paper, we have proposed a new computational approach that overcomes these limitations since it decomposes the  $q$-dimensional integral into the sum of $1, 2, \ldots , s$-dimensional integrals, being  $s < q$, that can be easily approximated using classical Gauss-Hermite quadrature techniques.  We have shown that, even if  DRM uses less information than  AGH, the DRM estimators share the same asymptotic properties of the AGH ones.  This is due to the fact that DRM is based on an  expansion of the integrand, known as Cut-HDMR, that provides a better approximation than any truncated Taylor series expansion that contains a finite number of terms, generally of first and second order. In this regard, the DRM-based estimators are more accurate than the classical Laplace ones, that corresponds to the case of $s$ equal to 0. On the other hand,  we have shown that, in finite samples,  DRM and AGH produce similar bias in the estimates, but the former tends to underestimate the Monte Carlo variance. These discrepancies are more evident when $s=0$, and reduce as  $s$ increases.  

\begin{table}
\caption{\label{tab06} Estimated model parameters under DRM5 with standard errors in brackets for the non-stationary model, DRM5, s=3, ADAMS data set}\tiny \singlespacing
\begin{tabular}{lll} \hline
& $\widehat{\alpha}$ & $\widehat{\sigma}^2_{u_i}$\\ \hline
$SER71$          & 1.00         & 0.57 (2.61) \\
$SER72$          & 0.59 (0.30)  & 1.37 (0.89) \\
$SER73$          & 0.53 (0.34)  & 0.74 (0.89) \\
$SER74$          & 0.62 (0.41)  & 0.70 (1.06) \\      	
$SER75$          & 0.72 (0.31)  & 0.40 (1.23) \\
\hline
\end{tabular}
\end{table}

\begin{center}\begin{figure}\centering
\psfrag{s}{\tiny$s$}
\psfrag{na}{\tiny $n_{q}=5$}\psfrag{ne}{\tiny $n_{q}=7$}\psfrag{ni}{\tiny $n_{q}=11$}\psfrag{1e-10}{\tiny $1e-10$}\psfrag{1e-07}{\tiny $1e-07$}\psfrag{1e-04}{\tiny $1e-04$}\psfrag{1e-01}{\tiny $1e-01$}
\psfrag{0}{\tiny $0$}\psfrag{2}{\tiny $2$}\psfrag{4}{\tiny $4$}\psfrag{6}{\tiny $6$}\psfrag{8}{\tiny $8$}
\includegraphics[height=7cm,width=7cm]{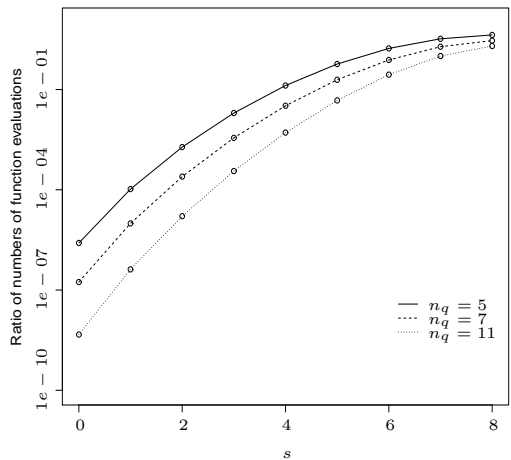}
\caption{Computational load, $q=8$ latent factors, $nq=5,7,11$ quadrature points.}\label{fig03}
\end{figure}\end{center}

The advantages of  DRM with respect to AGH are particularly evident from a computational point of view. When evaluating eq. (\ref{marg}) using the AGH quadrature rule with $n_{q}$  points in each dimension, the total number of function or response evaluations is $n_{q}^{q}$. In contrast, $\sum_{i=0}^{s}\left(\begin{array}{c}q\\s-i\end{array}\right)n_{q}^{s-i}$ function evaluations are required using the dimension reduction method. Figure \ref{fig03} shows how the ratio of these two function evaluation numbers varies with respect to $s$ for $q$ equal to 8 and $n_{q}$ equal to 5, 7  and 11. A reduction of the computational effort is achieved when the ratio $\sum_{i=0}^{s}\left(\begin{array}{c}q\\s-i\\\end{array}\right)n_{q}^{s-i}\left/n_{q}^{q}\right. < 1$. It can be noticed that the amount of reduction depends on both $s$ and $n_{q}$. In particular, for the univariate dimension reduction method the ratio has a magnitude of order $10^{-4}$, $10^{-5}$, and $10^{-7}$ when $n_{q} = 5, 7$ and 11, respectively. In the case of the bivariate DRM, the magnitude of the ratio is of order $10^{-3}$, $10^{-4}$, and $10^{-11}$ when $n_{q} = 5, 7$ and 11, respectively. Furthermore, even if not shown here, it is evident that the reduction is dramatically enhanced as $q$ increases.

A peculiar point of DRM technique is the choice of the number of terms involved in the computation. At this regard, we follow the same rule of thumb as for the choice of the number of quadrature points, that is we increase $s$ until  estimates stabilize. We underline that once the value of $s$ is selected, we can further improve the accuracy of the estimates   by increasing the number of quadrature points.

The main limitation of the proposed approach is its dependence on the likelihood function of the selected model. Even if the theoretical framework considered in this paper is general, the implementation of DRM on different models requires  ad hoc solutions since at the moment no commercial software is available yet. In this respect, we implemented DRM for GLLVMs in  R, and the code is available from the authors upon request.
 \vspace{-1.5cm}
\singlespacing
\bibliographystyle{chicago}
\bibliography{biblio}

\end{document}